\documentclass[%
 reprint,
 superscriptaddress,
nofootinbib,
 amsmath,amssymb,
 aps,
 prd
]{revtex4-2}
\usepackage[caption=false]{subfig}
\usepackage{physics}
\usepackage{bm}
\usepackage{graphicx}
\usepackage{dcolumn}
\usepackage{bm}
\usepackage{tikz}
\usepackage{adjustbox}
\definecolor{orcidlogocol}{HTML}{A6CE39}
\newcommand{\mathbit}[1]{\bm{#1}}
\usepackage{orcidlink}
\usepackage{hyperref}

\makeatletter
\renewcommand{\p@subfigure}{\thefigure}

\makeatother
\begin{document}


\title{Stone Skipping Black Holes in Ultralight Dark Matter Solitons}

\author{Alan Zhang\,\orcidlink{0009-0000-1187-0755}} 
 \email{Alan.Zhang1@anu.edu.au}
 \email{Zhana142@McMaster.ca}
 \affiliation{
 Center for Gravitational Astrophysics,Research School of Physics
\& Research School of Astronomy and Astrophysics, Australian National University, Canberra ACT 2601, Australia
}%


\author{Yourong Wang\,\orcidlink{0000-0003-0340-0758}}
 \email{yourong.wang@uni-goettingen.de}
 \affiliation{
 Institut für Astrophysik, Georg-August-Universität Göttingen, D-37077 Göttingen, Germany
}%

\author{J. Luna Zagorac\,\orcidlink{0000-0003-4504-1677}}
 \email{luna.zagorac@mcgill.ca}
 \affiliation{Perimeter Institute for Theoretical Physics, 31 Caroline St. N, Waterloo, ON N2L2Y5, Canada}
 \affiliation{Department of Physics \& Trottier Space Institute, McGill University, Montréal, QC H3A 2T8, Canada}

\author{Richard Easther\,\orcidlink{0000-0002-7233-665X}}
 \email{r.easther@auckland.ac.nz}
 \affiliation{Department of Physics, University of Auckland, Private Bag 92019, Auckland, New Zealand}

\date{\today}

\begin{abstract}
The orbit of a black hole moving within an ultralight dark matter (ULDM) soliton is naively expected to decay due to dynamical friction. However, in isolated near-circular soliton--black-hole systems, single black holes can undergo ``stone skipping'', with their orbital radius varying quasi-periodically. We show that, within this controlled setting, stone skipping is driven by a dipole excitation of the soliton. We model the effect as a resonance in a forced, damped harmonic oscillator, demonstrating that the coherent response of the soliton can significantly modify the dynamics of objects orbiting within it. In this regime, a dipole perturbation of a soliton can modify inspiral timescales when the black hole masses are significantly smaller than the soliton mass, with implications for supermassive black hole dynamics, the final parsec problem and gravitational wave observations in a ULDM cosmology.
\end{abstract}
\maketitle
\section{Introduction}
\label{sec:I}

Identifying the nature of dark matter  is a key  challenge for astrophysics and fundamental science. On cosmological scales,  Cold Dark Matter (CDM), within the $\Lambda$CDM paradigm, successfully describes the formation of large-scale structure and the anisotropies in the cosmic microwave background~\cite{planck2020, springel2005, frenk2012, primack2012}. However,  tensions at galactic and sub-galactic scales motivate candidates with more complicated dynamics than pure CDM. These small-scale challenges include the cusp-core~\cite{moore1994, deblok2010, oh2011},  missing satellites and the too-big-to-fail problems~\cite{bullock2017, weinberg2015, boylankolchin2011}. Baryonic feedback mechanisms may alleviate these discrepancies but models that naturally suppress small-scale power continue to be interesting~\cite{spergel2000, feng2010, bertone2005}.

Ultralight Dark Matter (ULDM), also known as Fuzzy Dark Matter, is one such scenario~\cite{hu2000, marsh2016, hui2017, ferreira2021,  eberhardt2025review}. In these models dark matter consists of extremely light bosonic particles with masses in the range $m \sim 10^{-23}\text{--}10^{-19}\,\mathrm{eV}$ and de Broglie wavelengths up  to kiloparsec scales. Such ultralight scalars arise naturally in string theory where the compactification of extra dimensions generically yields a plenitude of axion-like particles~\cite{arvanitaki2010, svrcek2006, witten1984}.

The  astrophysical behaviour of ULDM is governed by the Schr\"odinger-Poisson equation~\cite{widrow1993, chavanis2011}. A key prediction of this framework is the presence of solitons, stable, self-gravitating ground-state configurations, at the centers of collapsed halos~\cite{schive2014, schive2014a, mocz2017, schwabe2016}. These solitons are supported by quantum pressure and are surrounded by a halo of fluctuating  granules which exhibit wave-like mutual interference~\cite{veltmaat2018, lin2018}. Observational constraints from the Lyman-$\alpha$ forest typically imply $m \gtrsim  10^{-21}\,\mathrm{eV}$~\cite{irsic2017, rogers2021, armengaud2017}. Stellar-dynamical heating constraints from the smallest ultra-faint dwarfs can yield substantially stronger limits, $m \gtrsim 10^{-19}\, \mathrm{eV}$, with the precise value sensitive to modeling assumptions~\cite{dalal2022, may2025, eberhardt2025}.

A key arena for ULDM dynamics is the interaction between solitonic cores and the supermassive black holes (SMBHs) that appear to be present at the centers of all large galaxies. A massive object moving through a background medium experiences dynamical friction, a drag force caused by the object's gravitational wake~\cite{chandrasekhar1943, chandrasekhar1943a, binneytremaine2008}. In ULDM environments, however, the wave-mechanical nature of the medium introduces qualitatively new effects~\cite{hui2017, lancaster2020, baror2019}. Unlike collisionless or gaseous backgrounds~\cite{ostriker1999, just2011}, the coherent response of the condensate can produce oscillatory wakes and feedback loops which are not captured by Chandrasekhar-type formulae~\cite{edwards2018, wang2022, boey2024, vicente2022, cardoso2022}. Specifically, Wang and Easther~\cite{wang2022} demonstrated that a black hole orbiting inside a soliton excites coherent modes that backreact on the trajectory. This leads to ``stone skipping,'' a non-monotonic secular evolution of the orbital radius. Boey et al.~\cite{boey2024} further confirmed that soliton backreaction induces ``reheating'' that transfers kinetic energy back to the black hole.

The implications of stone skipping for the evolution of SMBH binaries are therefore model-dependent. Equal-mass SMBH binaries in initially unperturbed solitons do not excite the required dipole at leading order and do not show stone skipping in the simulations of Ref.~\cite{boey2024}; in very massive galaxies ULDM may damp orbital motion to the extent that gravitational wave emission in the pulsar timing band~\cite{nanograv2023, afzal2023, epta2023, ppta2023} is suppressed~\cite{Tiruvaskar:2025hxl}. More broadly, this has implications for the ``final parsec problem,'' which describes the tendency of SMBH binaries to stall at separations where gravitational-wave emission is inefficient~\cite{milosavljevic2003, begelman1980, kelley2017}. While mechanisms such as stellar hardening and gas torques have been proposed to bridge this gap~\cite{khan2013, vasiliev2015, gould2000}, ULDM and related dark-sector effects, including dynamical friction, have also been suggested as possible aids to binary hardening~\cite{koo2024,alonso-alvarez2024,boey2025,Barmak2024}. However, it also seems soliton backreaction can slow orbital decay in some circumstances. Consequently, it is well worth understanding the detailed dynamics of stone skipping, both for their intrinsic interest and their implications for SMBH dynamics. We use fully coupled Schr\"odinger-Poisson simulations to identify the coherent response of a perturbed soliton, and then use mode-filtered test-particle experiments  to diagnose which components of this response produce the observed orbital behaviour. In these reduced runs the black-hole backreaction on the wavefunction is  switched off and an empirical drag term supplies the  secular sink. 

We find that the large repeated rebounds characteristic of stone skipping are driven by the dipole sector: suppressing the dipole removes the effect, while adding it to backgrounds with no dipole recovers it. This identifies stone skipping as the wave-mechanical counterpart of a familiar response of self-gravitating systems: dipole, or seiche-like, oscillations that can remain weakly damped and exert a large-scale time-dependent force~\cite{weinberg1989, weinberg1994, weinberg2023, tremaineweinberg1984}. 

Related oscillations and random-walk behaviour of FDM soliton cores have been interpreted as interference between the ground state and excited states in a fixed potential~\cite{li2021}, while our eigenmode analysis follows the perturbative framework of Zagorac et al.~\cite{zagorac2022}. We  capture the same mechanism with a forced, damped oscillator: the coherent dipole supplies periodic driving, dynamical friction supplies dissipation, and stone skipping occurs when the forcing frequency lies near the natural epicyclic frequency so that energy transfer from the soliton can transiently overcome secular damping.

The structure of this paper is as follows. Section \ref{sec:II} introduces the ULDM-black hole system governed by the Schr\"odinger-Poisson equations and presents a modified formulation in which the black hole is treated as a test particle. Section \ref{sec:III} describes the eigenmode decomposition of the ULDM wave function in terms of expansion coefficients $c_{nlm}(t)$ and Section~\ref{sec:excited} describes their values in simulations with both single and binary black holes. In Section \ref{sec:IV} we treat the black hole as a test particle and systematically add excited modes to confirm that the dipole modes are the key ingredient for stone skipping in the reconstruction protocol. Section~\ref{sec:V} develops a semi-analytic model that interprets stone skipping as a classical forced, damped oscillator. Section~\ref{sec:VI} presents our conclusions.

\begin{figure*}[tb]
  \centering
  \includegraphics[width=\textwidth]{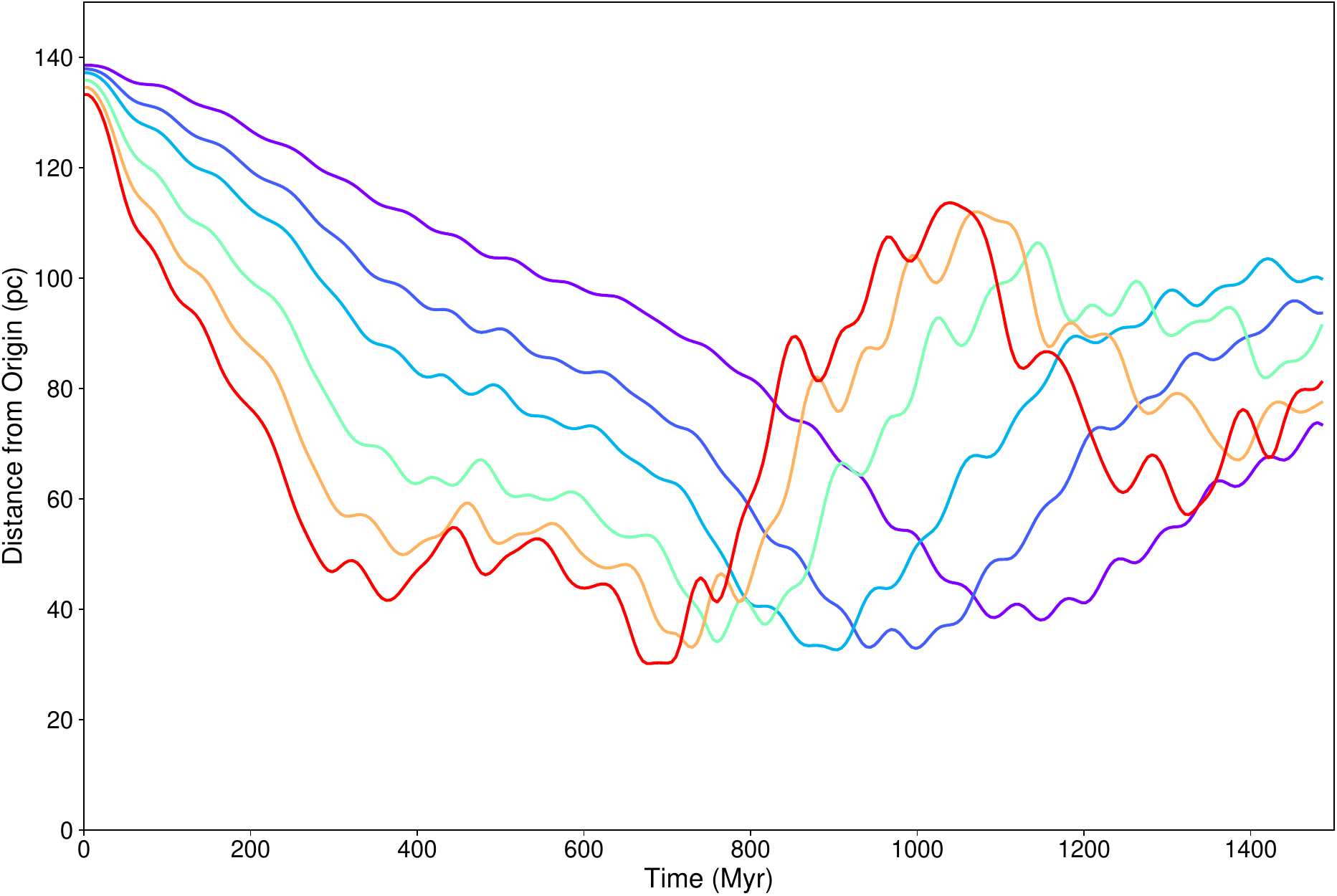}
  \caption{\label{fig:1} Time evolution of the orbital radius of black holes in initially circular orbits and masses $1$, $1.5$, $2$, $3$, $4$ and $5$\% of the soliton mass, from top to bottom at $t=0$. The initial orbital radii differ because, while the initial separation between the black hole and the soliton centroid is fixed at $140\,\mathrm{pc}$, more massive black holes lie closer to the system's center of mass. The physical radius of the simulation box is $1.25\,\mathrm{kpc}$.}
\end{figure*}

\section{ULDM Dynamics}
\label{sec:II}
ULDM can  be modeled as a dilute bosonic condensate or superfluid on macroscopic scales~\cite{edwards2018, ferreira2021, hui2017}. Its dynamics are  governed by the Schr\"odinger-Poisson system,
\begin{eqnarray}
\label{eq:sp}
i\hbar\frac{\partial\Psi}{\partial t} &=& \left[-\frac{\hbar^{2}}{2 m} \nabla^{2}+m\left(\Phi_U+\Phi_{E}\right)\right] \Psi, \\
\nabla^{2} \Phi_{U} &=& 4 \pi G m|\Psi|^{2},
\end{eqnarray}
where $\Psi(\mathbit{r},t)$ is the macroscopic wave function and $n(\mathbit{r},t)=|\Psi(\mathbit{r},t)|^2$ is the particle number density. The ULDM particle mass is taken to lie in the range
$m\sim 10^{-23}\text{--}10^{-19}\,\mathrm{eV}$~\cite{ferreira2021, marsh2016, hui2017}. The gravitational potentials $\Phi_U$ and $\Phi_{E}$ denote the ULDM gravitational potential and that of all other masses in the system, respectively.  In our case these are black holes, which we treat as Plummer spheres in our  numerical simulations, with potential
\begin{equation}
\mathrm{\Phi}_{BH}\left(r\right)=-G\frac{M_{BH}}{\sqrt{r^2+a^2}},
\label{eq:bh_p}
\end{equation}
where $a$ is the Plummer radius and $M_{BH}$ is the mass. The general equations of motion are
\begin{equation}
\ddot{\mathbit{r}}_{BH,i}
=-\nabla\mathrm{\Phi}_U\!\left(\mathbit{r}_{BH,i}\right)
-\sum_{j\neq i}\nabla\mathrm{\Phi}_{BH,j}\!\left(\mathbit{r}_{BH,i}\right) \, ,
\label{eq:bh1}
\end{equation}
but we are primarily interested in systems with one or two black holes.

Our focus is the interaction between black holes and the ULDM solitons which form naturally at the centers of ULDM halos ~\cite{schive2014,schive2014a}.  The soliton profile cannot be obtained in closed-form but is well approximated by~\cite{schive2014} 
\begin{equation}
\rho_s\left(r\right)\approx
\frac{1.9\left(m/{10}^{-23}\mathrm{eV}\right)^{-2}{(r_c/\mathrm{kpc})}^{-4}}
{\left[1+9.1\times{10}^{-2}{(r/r_c)}^2\right]^8}
\,M_\odot\,\mathrm{pc}^{-3} \, ,
\end{equation}
where $r_c$ is the core radius, defined by $\rho_s(r_c)=\rho_s(0)/2$. 

We employ both numerical simulations and expansions of the soliton in terms of its eigenmodes. Simulations are performed with the pseudospectral code \textsc{PyUltraLight}~\cite{edwards2018,wang2022}; the  expansion is developed in Ref.~\cite{zagorac2022}. Our starting configurations are a soliton and one or two black holes  in circular orbits about the overall center of mass. Unless stated otherwise, the simulations  employ a canonical set of parameters: ULDM particle mass of $10^{-21}$~eV and a soliton of total mass $M_{\rm tot}\approx 1.2\times 10^{7}\,M_\odot$, corresponding to a half-mass radius of $r_c\approx198.4 ~\mathrm{pc}$ evolved in a box of radius $r_{\max}=2500~\mathrm{pc}$ for a physical duration of $3000~\mathrm{Myr}$. The spatial resolution is $N=128$ with black hole(s) initialized on circular orbits at a radius of $140~\mathrm{pc}$. Varying the resolution and box size leaves the qualitative behavior unchanged, consistent with previous work based on staggered-leapfrog solutions to the Schr\"{o}dinger-Poisson system    \cite{edwards2018,zagorac2022,Zagorac:2022xic,wang2022,boey2024}. In particular, these analyses show that  while  the generalized Chandrasekhar dynamical friction \cite{hui2017} can differ from the instantaneous force,  it works very well (almost surprisingly so) on a time-averaged basis.


The na\"\i ve expectation is that dynamical friction \cite{chandrasekhar1943,chandrasekhar1943a,wang2022,boey2024} will reduce the orbital radius of the black hole so that it ``sinks'' to the center of the soliton, and this indeed happens initially. However, for stone skipping trajectories the inspiral reverses at some point and the orbital radius increases significantly and the cycle repeats. This behavior is clearly due to the black hole interacting with the now-oscillating soliton which is excited by the transfer of energy from the orbiting black hole. Stone skipping trajectories are illustrated in Figure~\ref{fig:1}.

Operationally, ``stone skipping'' denotes non-monotonic secular radial evolution in which the orbit undergoes repeated outward excursions after an initial decay. In practice this means that, after averaging over the short orbital-period oscillations, the coarse-grained radial velocity changes sign and the rebound amplitude is comparable to or larger than the local inspiral over the preceding interval. This is a qualitative description, but we expect to see a change in orbital radius of at least 25\% from the minimum value on a timescale at least several times larger than a single orbital period. 

The analysis is challenging because the black hole continuously excites the soliton as it moves within it. We separate these two aspects of the dynamics by simulating the motion of test particles in an excited soliton   with an empirical drag term, ``turning off'' the  backreaction of the black hole on the soliton.  The excited soliton has the general form 
\begin{equation}
\widetilde{\rho}_s\left(\mathbit{r}\right)=\rho_s\left(r\right)+\delta\rho\left(\mathbit{r}\right).
\end{equation}
where $\rho_s(r)$ is the ground state and  $\delta\rho(\mathbit{r},t)$ can be expressed via the eigenmode expansion.  The equation of motion  becomes 
\begin{equation}
    \begin{aligned}
\ddot{\mathbit{r}}_{BH,i}
=-\nabla\widetilde{\mathrm{\Phi}}_U\!\left(\mathbit{r}_{BH,i}\right)
&-\sum_{j\neq i}\nabla\mathrm{\Phi}_{BH,j}\!\left(\mathbit{r}_{BH,i}\right)
\\&+\frac{1}{M_{BH}}\mathbit{f}\left(\mathbit{r}_{BH,i},t\right),
\label{eq:bh2}
\end{aligned}    
\end{equation}
where $\widetilde{\Phi}_U$ is the ULDM self-potential evolved with $\Phi_{E}=0$, to distinguish it from $\Phi_U$ in Eq.~\eqref{eq:bh1}, and $\mathbit{f}(\mathbit{r}_{BH},t)$ is an empirical drag force. For the reconstructed tests we use
%
\begin{align}
 \mathbit{f}\left(\mathbit{r}_{BH},t\right)
&=-4\pi G^2 A  \rho\left(\mathbit{r}_{BH},t\right) \times \nonumber \\ & \quad  \frac{M_{BH}^2m^2\left|\mathbit{r}_{BH}\right|^2}{3\hbar^2}
\frac{\mathbit{v}_{\mathbit{rel}}}{\left|\mathbit{v}_{\mathbit{rel}}\right|} \, ,
\label{eq:df}
\end{align}
where $\rho(\mathbit{r}_{BH},t)$ is the ULDM density evaluated at the BH position $\mathbit{r}_{BH}$ at time $t$ and $\mathbit{v}_{\mathbit{rel}}=\mathbit{v}_{\mathbit{BH}}-\mathbit{v}_{\mathbit{ULDM}}$ is the black hole velocity relative to the ULDM flow field. Eq.~\eqref{eq:df} should be read as an effective version of the wave-mechanical ULDM dynamical-friction estimate, not as a Chandrasekhar formula transplanted without modification. 

In the uniform-background calculation of Hui et al.~\cite{hui2017}, as written in the notation of Wang and Easther~\cite{wang2022}, the force can be expressed as $F_{\rm DF}=4\pi\rho\,\overline C(\widetilde b)(GM_{\rm BH}/v_{\rm rel})^2$, where $\widetilde b$ is the travelled-distance cutoff in de Broglie units. For $\widetilde b\ll1$, $\overline C\simeq \widetilde b^2/3$. Taking the local wake scale to be of order the orbital radius gives the scaling in Eq.~\eqref{eq:df}, while the dimensionless coefficient $A$ absorbs the remaining cutoff, geometry, and finite-soliton effects~\cite{hui2017,wang2022,lancaster2020}. Such order-unity estimates give the correct scale for secular inspiral in controlled comparisons, but the fully coupled Schr\"odinger-Poisson response can also excite coherent soliton modes and make the effective friction history oscillatory~\cite{wang2022,boey2024}. 

We treat $A$ as providing a heuristic normalization, not as a universal calibration of ULDM drag. Because $\widetilde{\Phi}_U$ is evolved with $\Phi_E=0$ in our reconstructions the test particle does not create a  Schr\"odinger-Poisson wake in the  background; Eq.~\eqref{eq:df} supplies the secular sink that was  removed by switching off this backreaction -- it is not added on top of a simultaneously evolved wake generated by the same test particle. In the fully coupled simulations, by contrast, no empirical drag is inserted: the wake, coherent soliton response, and black-hole backreaction are evolved together. We have implemented these capabilities in a branch of \textsc{PyUltraLight}.\footnote{The \textsc{PyUltraLight} fork used in this work is available at \url{https://github.com/Ailun-Zhang/PyUL_SK}. The post-processing and visualization scripts used to analyse the \textsc{PyUltraLight} wavefunction outputs are available at \url{https://github.com/Ailun-Zhang/ULDM-Eigenmode-Toolkit}.}


\section{Eigenmode Decomposition}
\label{sec:III}
\begin{figure*}[t]
  \centering
      \includegraphics[width=\textwidth]{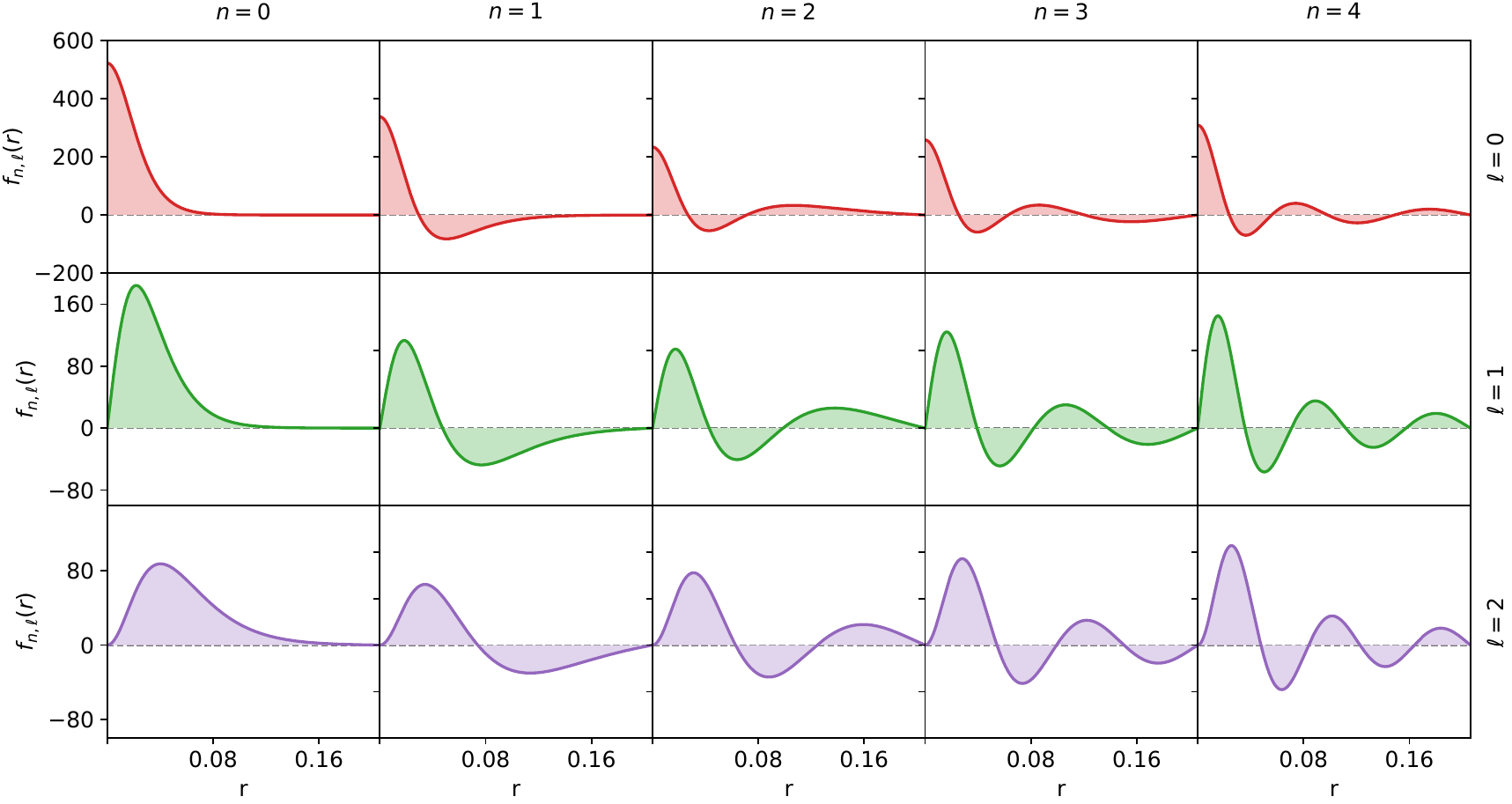}
  \caption{\label{fig:2} ULDM radial eigenfunctions with $n\le 4$ and $\ell\le 2$. The quantum number $n$ sets the number of radial nodes, while $\ell$ determines the small-$r$ asymptotic behavior, $f_{n\ell}(r)\propto r^\ell$ as $r\to 0$. Modes with the same $\ell$ are shown using the same color ($\ell=0$ in red, $\ell=1$ in green, and $\ell=2$  in purple); the same convention is adopted in subsequent figures.}
\end{figure*}

Our approach is based on Ref.~\cite{zagorac2022}, which we generalize to include the full set of angular states. Related eigenmode analyses have also been used to interpret soliton-core oscillations in FDM halos~\cite{li2021}. The macroscopic ULDM wave function is equivalent to an effective single-particle wave function up to its normalization so we may construct eigenmodes using standard quantum-mechanical methods. In operator form the Schr\"odinger sector of the system is
\begin{equation}
i\hbar\,\frac{\partial}{\partial t}\ket{\Psi}=\hat{H}(t)\ket{\Psi},
\qquad
\hat{H}(t)=\hat{H}_0+\hat{V}(t),
\end{equation}
where we separate a time-independent part from a time-dependent perturbation. The unperturbed Hamiltonian $\hat{H}_0$ consists of the usual kinetic term and the gravitational potential $\Phi_0$ of the ground-state soliton
\begin{equation}
\hat{H}_0=-\frac{1}{2}\nabla^2+\Phi_0.
\label{eq:H0}
\end{equation}
The eigenstates are defined via the time-independent Schr\"odinger equation,
\begin{equation}
\hat{H}_0\ket{n,\ell,m}=E_{n\ell}\ket{n,\ell,m},
\label{eq:tise}
\end{equation}
where $(n,\ell,m)$ are the usual quantum numbers.  Eq.~\eqref{eq:tise} is spherically symmetric so, as usual, separation of variables yields
\begin{equation}
\psi_{n\ell m}(\mathbit{r})
=\braket{\mathbit{r}}{n,\ell,m}
=f_{n\ell}(r)\,Y_{\ell m}(\theta,\phi),
\label{eq:eigsta}
\end{equation}
where $Y_{\ell m}$ are the orthonormal spherical harmonics. The radial functions have no simple closed form but satisfy
\begin{equation}
\begin{aligned}
\left[
-\frac{1}{2}\frac{d^2}{dr^2}
+\frac{\ell(\ell+1)}{2r^2}
+\Phi_0(r)
\right]&u_{n\ell}(r)
=E_{n\ell}\,u_{n\ell}(r),\\
&u_{n\ell}(r)\equiv r\,f_{n\ell}(r),
\label{eq:radial}
\end{aligned}
\end{equation}
with boundary conditions $u_{n\ell}(0)=u_{n\ell}(r_{\max})=0$. Figure~\ref{fig:2} shows examples of the radial eigenfunctions for $n\le 4$ and $\ell\le 2$.

As described in Appendix~\ref{apx:A}, we must solve the radial problem numerically and the resulting eigenvectors are chosen to be orthonormal and complete in the discretized space. The eigenstates satisfy
\begin{equation}
    \braket{n',\ell',m'}{n,\ell,m}
=\delta_{nn'}\,\delta_{\ell\ell'}\,\delta_{mm'},\label{eq:ornml}
\end{equation}
\begin{equation}
\sum_{n\ell m}\ket{n,\ell,m}\bra{n,\ell,m}
=\mathbb{I}.
\label{eq:compt}
\end{equation}
The eigenstates Eq.~\eqref{eq:tise} are constructed for the ground-state soliton background and are  not eigenstates of the full time-dependent Hamiltonian in Eq.~\eqref{eq:sp} with its nontrivial external potential. Nevertheless, because they form a complete basis in the discretized finite-domain Hilbert space used in the simulations, they can be used to expand the ULDM wave function at even after it is perturbed by a black hole, up to the numerical truncation described in Appendix~\ref{apx:A}. We introduce an effective single-particle wave function
\begin{equation}
\psi(\mathbit{r},t)=\frac{1}{\sqrt{N}}\,\Psi(\mathbit{r},t),
\qquad
N=\int d^3\mathbit{r}\,|\Psi|^2,
\label{eq:sgwf}
\end{equation}
since the eigenstates \eqref{eq:eigsta} are normalized in the single-particle sense, whereas the macroscopic field $\Psi$ carries a different normalization and statistical meaning.\footnote{The macroscopic wave function output by \textsc{PyUltraLight} includes a factor of $\sqrt{m}$, and is related to the field appearing in Eq.~\eqref{eq:sp} by $\Psi_{\rm PyUL}=\sqrt{m}\,\Psi$. In the ``code units'' $c=\hbar=G=m=1$ this distinction disappears. }
We  expand the macroscopic ULDM  wave function 
\begin{equation}
\Psi(\mathbit{r},t)
=\sqrt{M_{\rm tot}}
\sum_{n\ell m} c_{n\ell m}(t)\,\psi_{n\ell m}(\mathbit{r}),
\label{eq:decomp}
\end{equation}
where $c_{n\ell m}(t)$ are complex, time-dependent coefficients and $M_{\rm tot}=Nm$ is the total soliton mass. The ground state $\ket{0,0,0}$ corresponds to the soliton itself. In the small-perturber regime (which applies here with black hole masses at the few-percent level) the ULDM field is a combination of the ground state and an admixture of excited modes whose amplitudes are much smaller than the ground state contribution. 

States with $\ell>0$ are aspherical, while those with $\ell=0$ and $n\neq 0$ are purely radial.  In analogy with multipole terminology, we further define, for each  $n$,
\begin{align}
\textit{Monopole:}\quad & c_{000}\,\psi_{000},
\\
\textit{Dipole:}\quad & \sum_{m=-1}^{1} c_{n1m}\,\psi_{n1m},
\\
\textit{Quadrupole:}\quad & \sum_{m=-2}^{2} c_{n2m}\,\psi_{n2m}.
\end{align}

The mode coefficients should therefore be interpreted as finite-domain diagnostics rather than exact continuum observables. The simulation box, radial boundary condition, finite grid, and truncation of the displayed mode set can all move small amounts of power between nearby modes or into a numerical noise floor. However, we can be confident in our overall results because the relevant mode amplitudes are much larger than the noise floor. 

\section{Soliton Excitations From Stone Skipping}
\label{sec:excited}
\begin{figure}
  \centering
  \includegraphics[width=\columnwidth]{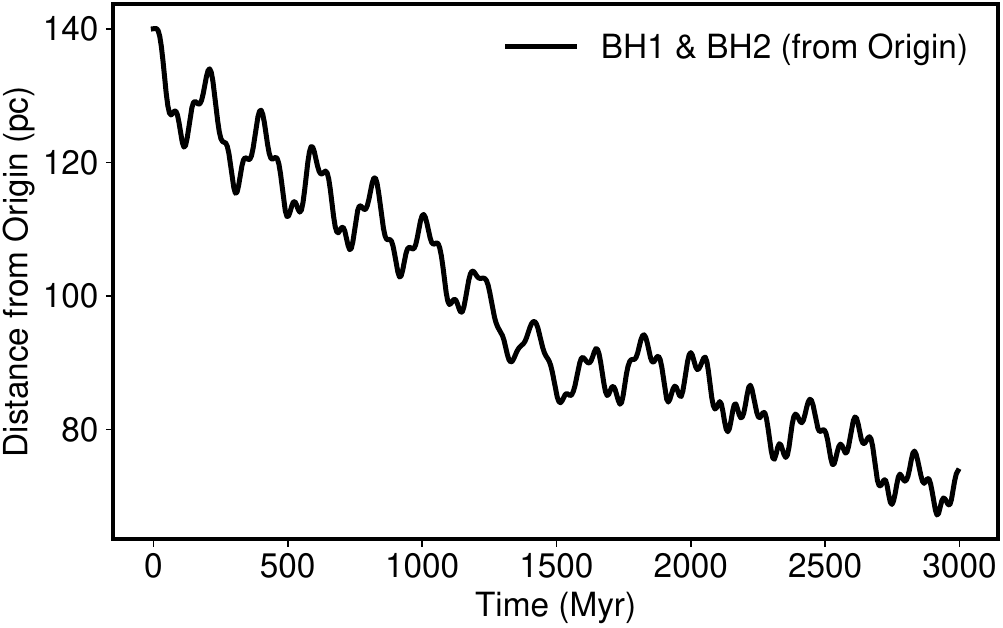}
  \caption{\label{fig:3} Evolution of the orbital radii of an equal-mass black hole binary, with each black hole having mass $2\%$ of the soliton mass.}
\end{figure} 

\begin{figure*}
  \centering
  \includegraphics[width=0.96\textwidth]{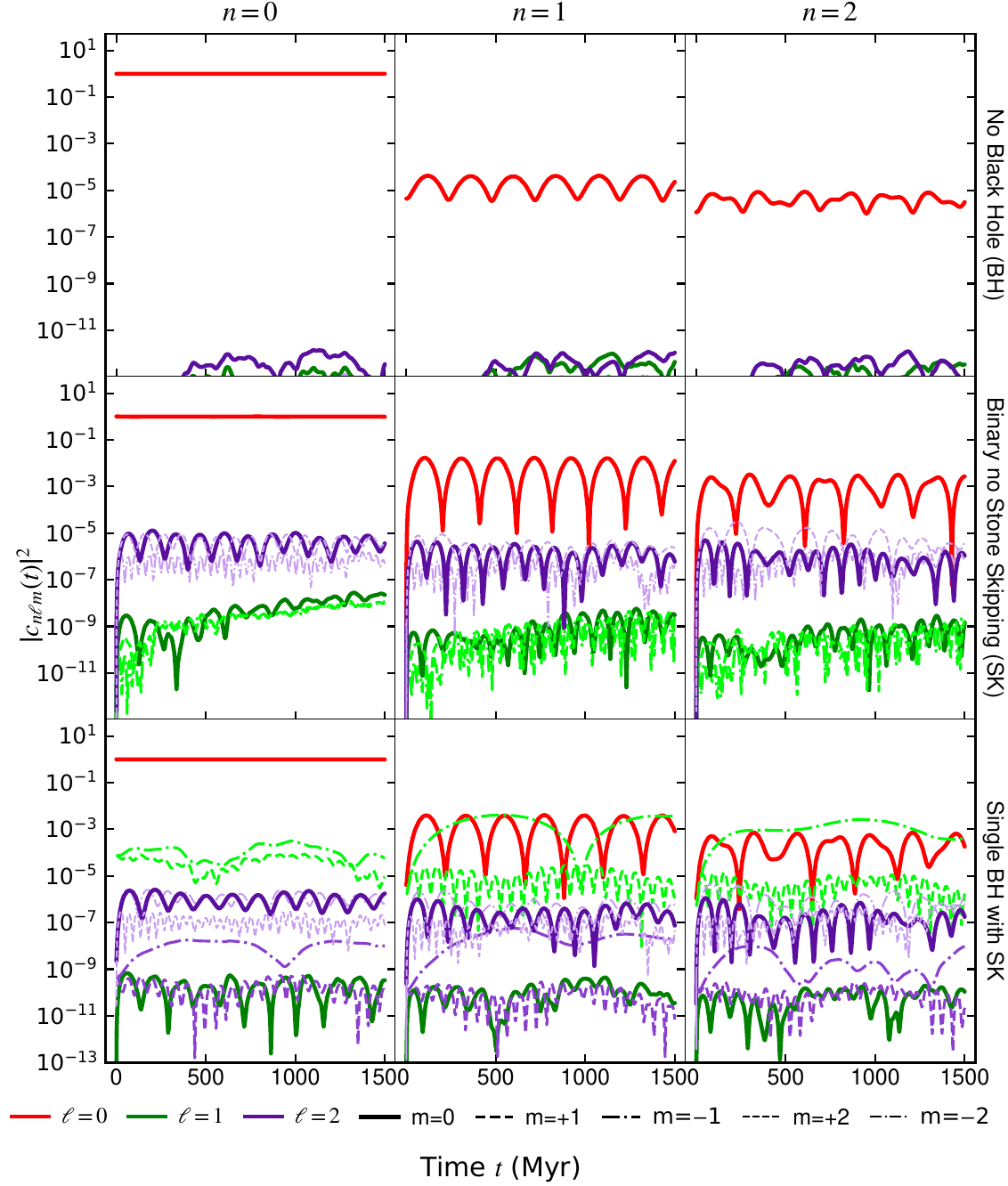}
  \caption{\label{fig:4} Evolution of $|c_{n\ell m}(t)|^2$ obtained from the eigenmode decomposition of the ULDM wave functions. Top row: pure soliton without a black hole. Middle row: soliton with an equal-mass binary with each black hole mass $2\%$ of the soliton mass; no stone skipping is observed. Bottom row: single black hole with mass $2\%$ of the soliton mass; stone skipping is observed. Colors denote mode groups: red for the monopole, green for the dipole, and purple for the quadrupole; within each group, shading lightens as $|m|$ increases. Line styles distinguish the azimuthal order $m$: thick solid curves for $m=0$, dashed for $m>0$, and dash-dotted for $m<0$, with line width decreasing as $|m|$ increases.}
\end{figure*}

As usual,  orthonormality~\eqref{eq:ornml} yields
\begin{equation}
c_{n\ell m}(t)=\frac{1}{\sqrt{M_{\rm tot}}}\int d^3\mathbit{r}\;
\Psi(\mathbit{r},t)\,\psi_{n\ell m}^{\ast}(\mathbit{r}).
\label{eq:cnlm}
\end{equation}
In this mean-field setting, $|c_{n\ell m}(t)|^2\lesssim 1$ can be interpreted as the fractional occupation (by mass/number) of the mode $\psi_{n\ell m}$ at time $t$. Since the black hole is much less massive than the soliton we expect the soliton to remain close to its ground state, {\em i.e.}\ $|c_{000}(t)|^2\simeq 1$ and the other coefficients to be much less than unity, and this is confirmed for a range of parameter choices. 

We decompose time-dependent ULDM wave functions obtained from simulations initialized with the pure-soliton profile for three configurations:
(i) a control run with no black hole, (ii) a binary where each black hole is $2\%$ of the soliton mass, and (iii) a single black hole that is $2\%$ of the soliton mass. The binary evolution is shown in Figure~\ref{fig:3}. Consistent with Refs.~\cite{boey2025,koo2024}, the black hole radii decrease steadily but not strictly monotonically, in contrast to the stone skipping shown in Figure~\ref{fig:1}.
\begin{figure*}[t]
  \centering
  \begin{minipage}[t]{0.49\textwidth}
    \centering
    \includegraphics[width=\linewidth]{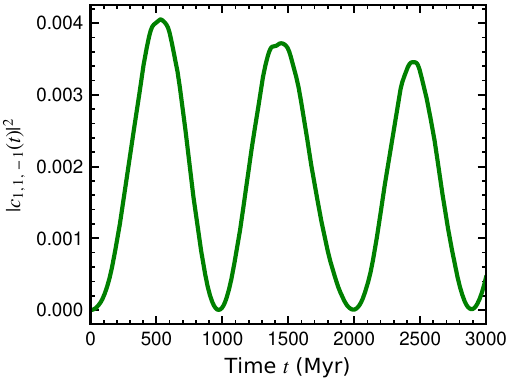}
    \caption{\label{fig:5} Evolution of the dipole component $|c_{1,1,-1}(t)|^2$ in the stone skipping run perturbed by a single black hole. The curve is well described by a $\cos^2$-type oscillation.}
  \end{minipage}\hfill
  \begin{minipage}[t]{0.49\textwidth}
    \centering
    \includegraphics[width=\linewidth]{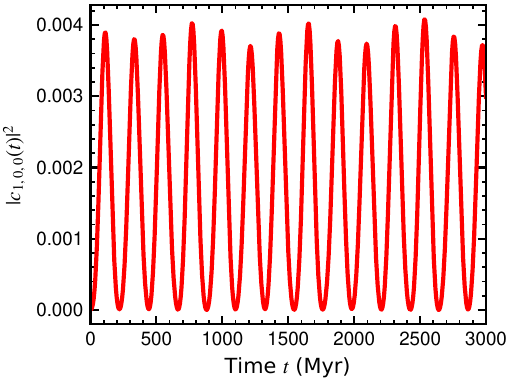}
    \caption{\label{fig:6} Evolution of the radial component $|c_{1,0,0}(t)|^2$ in the stone skipping run perturbed by a single black hole. The curve is approximately $\cos^2$-like.}
  \end{minipage}
\end{figure*}
\begin{figure*}
  \centering
  \subfloat[\label{fig:8a} Black hole trajectories with and without radial-mode excitations, without dipole modes.]{%
    \begin{minipage}[t]{0.475\textwidth}
      \centering
      \includegraphics[width=\linewidth]{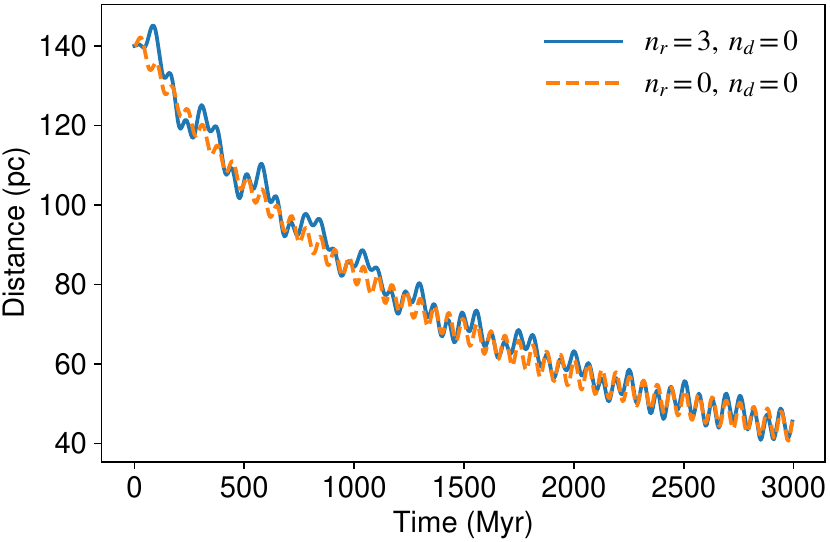}
    \end{minipage}%
  }\hfill
  \subfloat[\label{fig:8b} As panel~\ref{fig:8a}, but with dipole modes up to $c_{1lm}$.]{%
    \begin{minipage}[t]{0.475\textwidth}
      \centering
      \includegraphics[width=\linewidth]{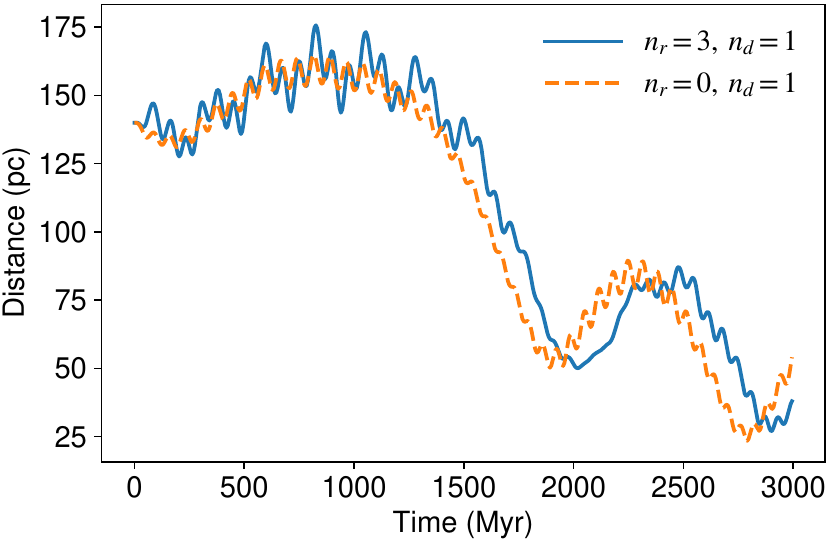}
    \end{minipage}%
  }

  \vspace{0.15em}

  \subfloat[\label{fig:8c} As panel~\ref{fig:8a}, but with dipole modes up to $c_{2lm}$.]{%
    \begin{minipage}[t]{0.475\textwidth}
      \centering
      \includegraphics[width=\linewidth]{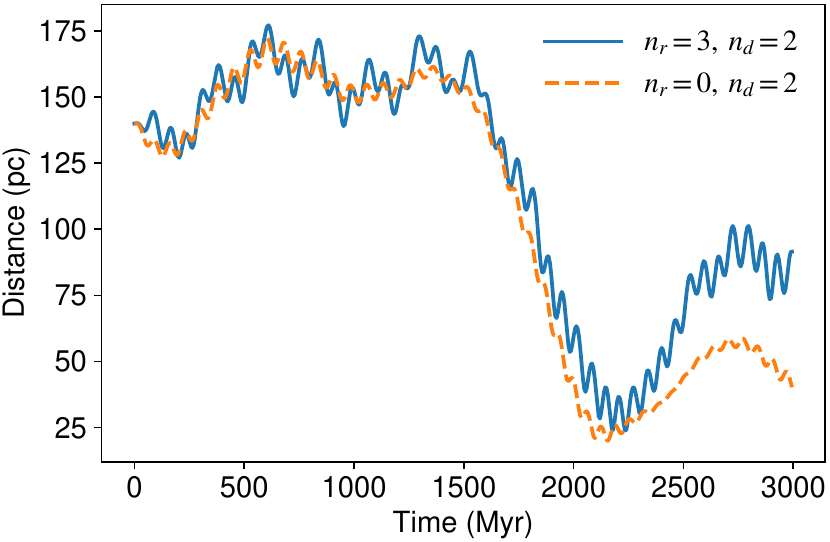}
    \end{minipage}%
  }\hfill
    \subfloat[\label{fig:8f} Black hole trajectories without radial modes.]{%
    \begin{minipage}[t]{0.475\textwidth}
      \centering
      \includegraphics[width=\linewidth]{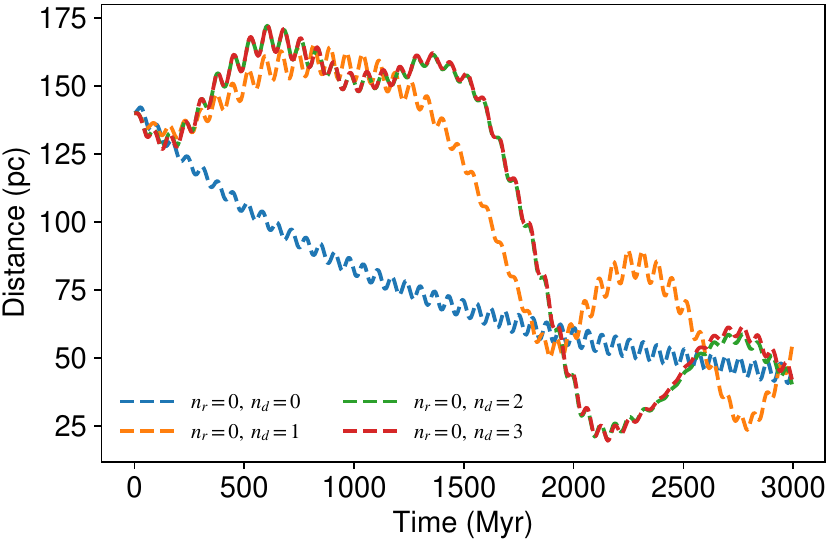}
    \end{minipage}%
  }
  \caption{\label{fig:8} Comparisons of black hole trajectories with (orange) and without (blue) radial-mode excitations at fixed dipole content; $n_r$ denotes the maximum radial quantum number and $n_d$ denotes the maximum $n$ included in the dipole modes.}
\end{figure*} 

\begin{figure*}[t]
  \centering
  \includegraphics[width=\textwidth]{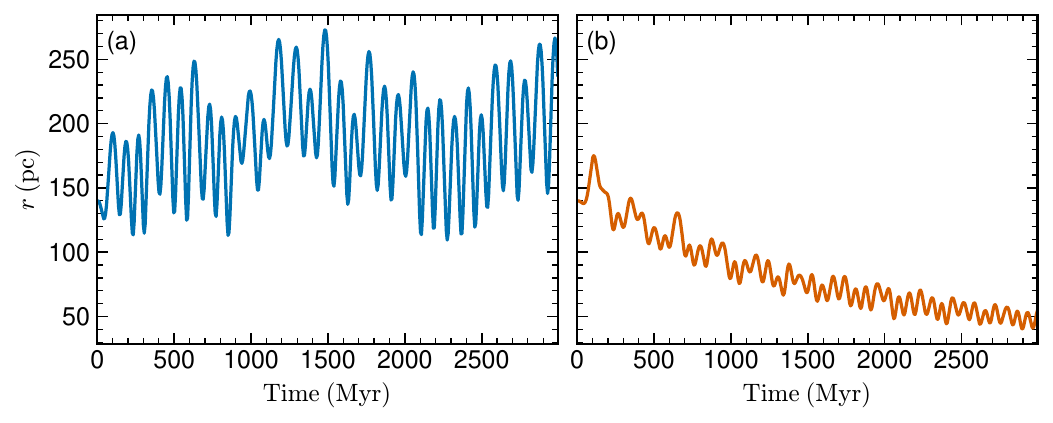}
  \caption{\label{fig:dipole_removed} Higher-mode test for the $2\%$ single-black-hole case. Both panels use a reconstructed background in which all modes with $n\le2$ and $\ell\le2$ are populated at a common amplitude, with phases extracted from the fully coupled simulation. Left: retaining the dipole sector produces large radial excursions. Right: removing all $\ell=1$ modes while retaining radial and quadrupole content removes the stone-skipping signal. This diagnostic reconstruction addresses the role of higher multipoles in the tested protocol, not all possible environments.}
\end{figure*}

The top row of Figure~\ref{fig:4} shows the decomposition of the pure soliton; as expected it is almost time independent. Aside from the monopole component (the soliton itself) there are small $\ell=0$ radial excitations with $n\neq 0$ which likely correspond to a slight mismatch in the initial profile and aspherical components remain below $\sim 10^{-12}$, consistent with numerical noise. The middle row shows the  equal-mass binary; the quadrupole is  excited, reaching a fractional contribution of $\sim 10^{-5}$ for $n=0$ and the dipole remains small. The bottom row is the single black hole case and both dipole and quadrupole components are appreciably excited with the dipole  dominant; for $n=1$, the weight rises to $\sim 0.004$. 

In particular, the mode $\ket{n=1,\ell=1,m=-1}$ is not only the largest  aspherical component but is roughly periodic in time.\footnote{The asymmetry between the $m=\pm1$ modes reflects the orbital direction of the black hole; if we reverse the direction the $m=1$ mode is larger.} We plot it separately on linear axes in Figure~\ref{fig:5} and can describe it phenomenologically via
\begin{equation}
|c_{1,1,-1}(t)|^2 \sim C\,\cos^2\!\left(\omega_{\rm dip}\,t+\phi_0\right),
\label{eq:dipfit}
\end{equation}
where $C$ sets the amplitude, $\omega_{\rm dip}$ is the dipole oscillation frequency, and $\phi_0$ is an initial phase.\footnote{The overall amplitude decays if the simulation is run for very long times. However, the decay rate decreases with increasing resolution, so this appears to be a numerical artifact, at least in part.} This oscillation can be interpreted as the effective driving force in a forced, damped oscillator model for stone skipping. To specify the driving term more fully we must analyze the frequency content of the \emph{complex} coefficient $c_{1,1,-1}(t)$ (without taking the modulus) and infer a plausible functional form from its spectrum. 

The very small coefficients in Figure~\ref{fig:4}, especially those in the pure-soliton and symmetry-forbidden sectors define an empirical numerical floor set by the finite grid, finite box, projection truncation, and imperfect cancellation of symmetries. We therefore use the coefficient hierarchy to identify robust, order-of-magnitude differences between runs and do not attach significance to features near the noise floor.  Additionally, the coefficients are continuously modified by the interaction between the soliton and the black hole potential -- if these were pure eigenstates we would expect the amplitudes to be roughly constant.  Some spherically symmetric radial modes ($\ell=0$, $n\neq 0$) can attain amplitudes comparable to the dipole modes.  Figure~\ref{fig:6} shows that these radial components exhibit oscillations with amplitudes comparable to those of the dipole modes, but with a higher frequency.  

While radial modes are excited in both single and binary black hole scenarios, significant dipole excitation is unique to the single black hole, suggesting that this is the primary driver of stone skipping. We now test this hypothesis  by evolving black holes in  backgrounds in which radial or dipole modes are  excited ``by hand''.

\section{Eigenmodes and Stone Skipping}
\label{sec:IV}

A sharper test of the correlation between ULDM excitations and stone skipping is to construct initial ULDM states containing specific eigenmodes and to evolve the black hole via Eq.~\eqref{eq:bh2} with the empirical drag Eq.~\eqref{eq:df}. Specifically, 
\begin{equation}
\Psi^{\vec{v}_0}_{\rm init}
=\sqrt{M_{\rm tot}} \sum_{n=0}^{N} \sum_{\ell=0}^{L} \sum_{m=-\ell}^{\ell} c_{nlm}(t_0)\ket{n,\ell,m}
\label{eq:iniPsi}
\end{equation}
for some (typically small) values of $N$ and $L$, with $c_{000}$ much larger than the other coefficients. We ensure that $M_{\rm tot}$ is actually the total mass by rescaling the ``raw'' monopole term $\bar c_{000}$
\begin{equation}
c_{000}
=\frac{\bar c_{000}}{|\bar c_{000}|}
\sqrt{1-\sum_{n,\ell,m\ne0}\left|c_{n\ell m}\right|^2}\, .
\label{eq:c000renorm}
\end{equation}
The superscript $\vec{v}_0$ in Eq.~\eqref{eq:iniPsi} indicates that the    configuration generally carries a bulk (center-of-mass) velocity,
\begin{equation}
\vec{v}_0
=\frac{\hbar}{m}\,
\frac{\int d^3r\;
\Im\!\left[\left(\Psi^{\vec{v}_0}_{\rm init}\right)^\ast
\nabla\Psi^{\vec{v}_0}_{\rm init}\right]}
{\int d^3r\;\left|\Psi^{\vec{v}_0}_{\rm init}\right|^2}\, .
\label{eq:v0}
\end{equation}
A Galilean boost 
\begin{equation}
\Psi^{0}_{\rm init}
=\exp\!\left(-\frac{i}{\hbar}m\,\vec{v}_0\cdot\vec{r}\right)\Psi^{\vec{v}_0}_{\rm init}.
\label{eq:boost}
\end{equation}
thus keeps the ULDM center of mass at the origin. 

We construct $\Psi^{0}_{\rm init}$ from subsets of the $c_{nlm}$ to isolate the terms responsible for stone skipping. We include radial modes ($\ell=0$) up to $n_r$ and dipole modes $c_{n1m}$ with $n\le n_d$. For definiteness we fix $t_0=2,500$~Myr and set the $c_{nlm}$ accordingly. The drag coefficient is fixed to $A=0.75$ throughout these reconstructions, a value that produces a secular decay comparable to the full run and therefore allows like-for-like comparisons between mode choices.  Note too that these diagnostic simulations have approximately constant $|c_{nlm}|^2$; the $\cos^2$ modulation in the full simulations is induced by forcing from the black holes and is suppressed in this approximation. Crucially, this does not imply a static potential, as the time evolution is determined by the real part of the coefficients (see Eq.~\ref{eq:C21} in Appendix~\ref{apx:C}). Figure~\ref{fig:8} shows that stone skipping is recovered when the dipole modes are included, whereas radial modes alone do not produce it within this protocol.  

The stronger higher-mode test is to start from a deliberately broad, artificially populated mode set and then remove only the dipole sector. This addresses the possibility that radial or quadrupole modes appear unimportant merely because the black hole did not excite them strongly enough in the original run. In Figure~\ref{fig:dipole_removed}, all modes with $n\le2$ and $\ell\le2$ are populated at a common amplitude while preserving the phases extracted from the fully coupled simulation. When the dipole sector is retained the test particle undergoes large stone-skipping-like radial excursions. When all $\ell=1$ modes are removed while the radial and quadrupole sectors remain present at comparable amplitude, the trajectory returns to secular decay with only small oscillatory modulations. 

\begin{figure*}[t]
  \centering
  \includegraphics[width=0.98\textwidth]{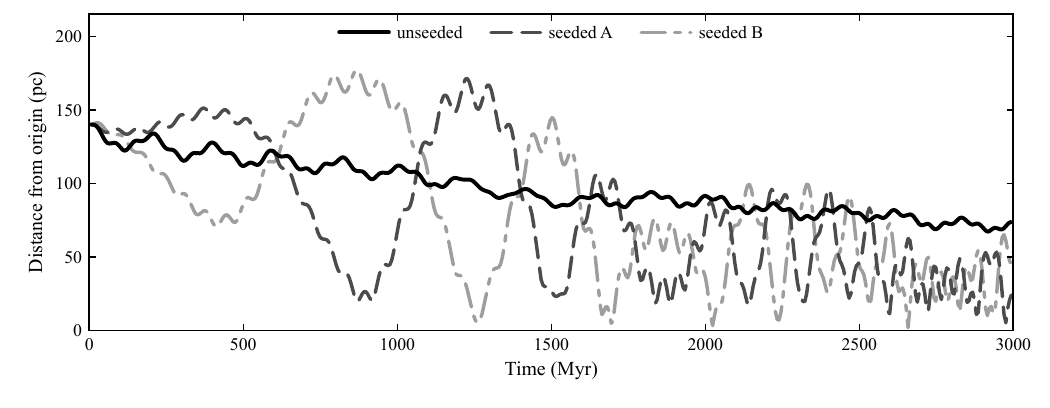}
  \caption{\label{fig:equal_mass_reduced_seeded} Reduced seeded-dipole diagnostic for an equal-mass binary, compared with the unseeded fully coupled baseline. The solid curve repeats the unseeded equal-mass binary in Figure~\ref{fig:3}; by symmetry the two black holes have the same orbital-radius evolution and only one curve is plotted. The dashed and dash-dotted curves show the two black holes in the reduced seeded-dipole diagnostic, where each black hole has mass $2\%$ of the soliton mass and is evolved in a prescribed soliton-plus-dipole background with the empirical drag term in Eq.~\eqref{eq:df}. Black-hole backreaction on the ULDM field is deliberately switched off for the reduced diagnostic.}
\end{figure*}

Before turning to a fully coupled solution, it is useful to separate the two questions that the reduced reconstructions can and cannot answer. Figure~\ref{fig:equal_mass_reduced_seeded} shows an equal-mass binary evolved in a soliton plus a seeded dipole background using the same reduced orbit equation and empirical drag prescription as the mode-isolation tests above -- the imposed dipole field can drive repeated radial rebounds. This is not a full dynamical demonstration: the ULDM background is prescribed, black-hole backreaction on that background is switched off, and the dissipative term is the effective prescription in Eq.~\eqref{eq:df}.

\begin{figure*}[t]
  \centering
  \includegraphics[width=0.98\textwidth]{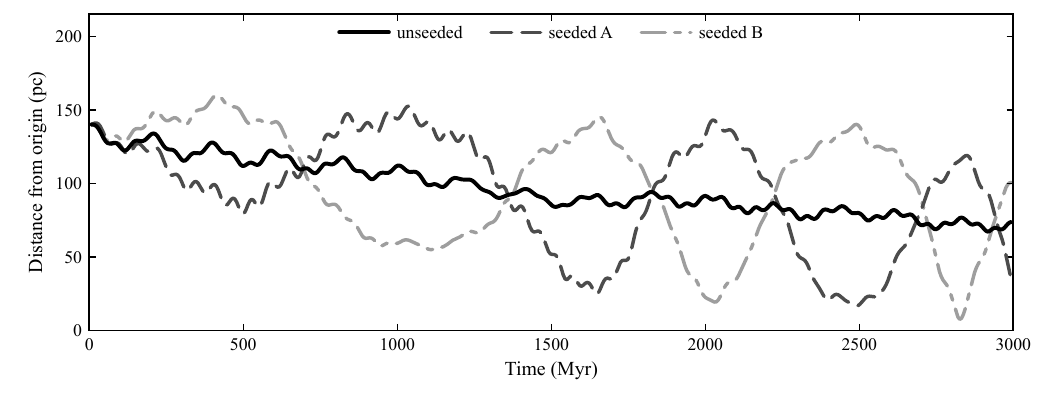}
  \caption{\label{fig:equal_mass_seeded_fullSP} Fully coupled seeded-dipole run for an equal-mass binary. The solid curve shows the unseeded equal-mass binary in Figure~\ref{fig:3}; by symmetry the two black holes have the same orbital-radius evolution and only one curve is plotted. The dashed and dash-dotted curves show the two black holes in a fully coupled evolution with the same binary mass scale but with a small initial $n=1$, $\ell=1$ dipole seed in the soliton. No empirical test-particle drag term is included in either simulation. The contrast shows that changing the initial dipole content alone can turn a symmetric inspiral into a stone-skipping trajectory in the fully coupled Schr\"odinger-Poisson system.}
\end{figure*}

The symmetry of the unseeded equal-mass binary makes the corresponding fully coupled test particularly clean. In the middle row of Figure~\ref{fig:4}, the equal-mass binary excites the quadrupole sector but leaves the dipole sector strongly suppressed, and the orbital radii in Figure~\ref{fig:3} show no large repeated rebounds. Figure~\ref{fig:equal_mass_seeded_fullSP} then makes a one-change comparison: the solid curve is the same unseeded equal-mass binary, while the dashed and dash-dotted curves are the two black holes in a fully coupled run with the same binary mass scale but a small initial $n=1$, $\ell=1$ dipole seed. The seeded run displays repeated, order-tens-of-parsec rebounds over the simulated interval. Thus breaking the field symmetry by adding a low-lying dipole, rather than changing the binary masses or inserting a test-particle drag force, is sufficient to restore stone-skipping-like motion in the fully coupled comparison.

Together, Figures~\ref{fig:8}--\ref{fig:equal_mass_seeded_fullSP} support the same bounded conclusion from complementary directions. Mode-filtered reconstructions identify the dipole sector as the part of the resolved soliton response needed for the large rebounds; the higher-mode run shows that comparably excited radial and quadrupole sectors do not replace it; and the fully coupled equal-mass run shows that an initial dipole seed restores rebound-like motion without an added drag force. This is strong evidence for a dipole-mediated mechanism.

The fully coupled seeded run is especially important for the interpretation of Eq.~\eqref{eq:df}. The reduced reconstructions use that effective drag term to provide a  secular sink while the mode content is varied by hand, so they are diagnostic rather than self-contained dynamical simulations. Figure~\ref{fig:equal_mass_seeded_fullSP} removes this ambiguity in identifying the physical mechanism: both the unseeded and seeded equal-mass binaries are evolved with the same Schr\"odinger-Poisson backreaction, and neither contains an inserted test-particle drag. The only intended change is the low-lying dipole content of the initial soliton. The fact that the unseeded binary decays without sustained rebounds whereas the seeded binary develops large repeated excursions is therefore difficult to attribute to the empirical damping prescription. It is instead the fully coupled counterpart of the mode-filtered statement that the dipole sector controls the rebound channel in this isolated near-circular setup.

\section{A semi-analytic resonance model}
\label{sec:V}

We now develop a semi-analytic model of stone skipping. In this picture, the time-dependent gravitational field induced by the evolution of the dipole excitation is an external driving force while the dynamical friction experienced by a black hole provides dissipation. The parameter $\epsilon$ denotes the amplitude of the selected excited mode, $\Phi_{1m}(r)$ denotes the corresponding dipole contribution to the gravitational potential, and $\gamma_r$ and $\gamma_\phi$ are phenomenological damping rates for the radial and azimuthal components of the reduced drag. The quantities $\Omega_0$ and $\kappa$ are, respectively, the circular-orbit and epicyclic frequencies at the reference radius $r_0$, while $\Gamma$ below is the standard oscillator damping parameter. The derivation assumes a near-circular orbit, a small dipole perturbation, and weak enough damping for a resonance window to be meaningful.

\begin{figure*}
  \centering
  \includegraphics[width=0.91\textwidth]{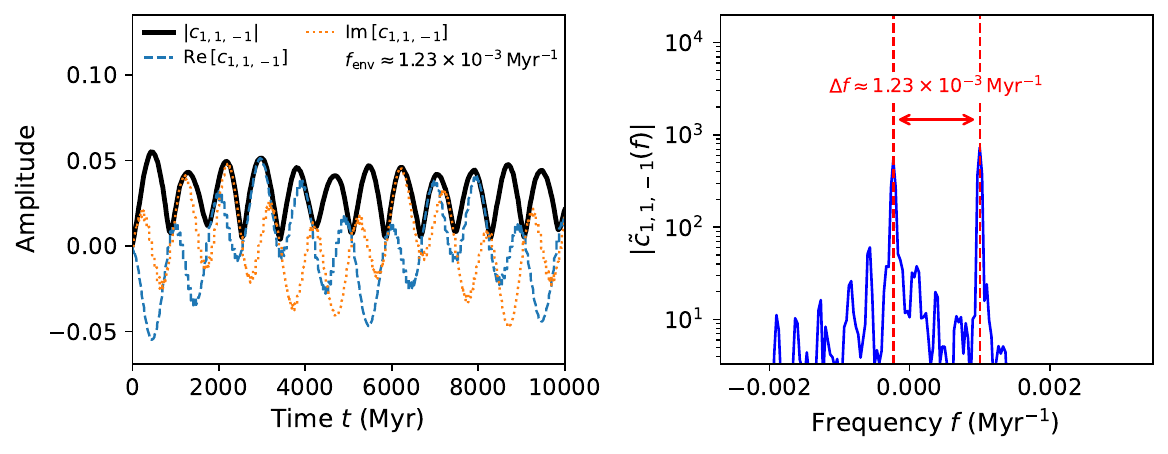}
  \caption{\label{fig:14} Spectral analysis of the coefficient $c_{1,1,-1}(t)$. Left: real and imaginary parts in the time domain. Right: FFT spectrum showing two strong peaks associated with an apparent splitting. The positive frequency separation is $\Delta f=|f_2-f_1|$; the corresponding beat angular frequency in Eq.~\eqref{eq:wdip} is $\omega_{\rm dip}=\pi\Delta f$. Additional negative-frequency peaks have amplitudes at the $\mathcal{O}(10^{-2})$ level and are neglected.}
\end{figure*}

Concretely, we write the equations of motion for a black hole confined to a two-dimensional plane in a background consisting of the soliton plus a single excited mode of amplitude $\epsilon$, and include the effective dynamical-friction prescription (Eq.~\eqref{eq:df}). In polar coordinates, the dynamical friction is taken to be proportional to two damping coefficients, $\gamma_r$ and $\gamma_\phi$, for the radial and azimuthal components, respectively. We then expand the equations about a circular orbit,
\begin{equation}
r(t)=r_0+\xi(t),\qquad \phi(t)=\Omega_0 t+\eta(t),
\end{equation}
where $(r,\phi)$ are the black hole coordinates, $r_0$ is the circular-orbit radius, $\Omega_0$ is the corresponding orbital frequency, and $\xi(t)$ and $\eta(t)$ are first-order perturbations about the circular solution. This procedure yields a linear system of differential equations for $(\xi,\eta)$.

We first consider the conservative limit in which both the excited mode and dynamical friction are switched off, {\em i.e.}\ $\epsilon=\gamma_r=\gamma_\phi=0$. After the change of variables $\dot{\xi}=u$, the linearized system reduces to a simple harmonic oscillator,
\begin{equation}
\ddot{u}+\kappa^2 u=0,
\label{eq:sho}
\end{equation}
where $\kappa$ is the epicyclic frequency of the soliton--black hole system. It is determined by the local soliton potential and the orbital frequency,
\begin{equation}
\kappa^2=\Phi_0''(r_0)+3\Omega_0^2.
\label{eq:kappa}
\end{equation}

If we turn on the excited mode while keeping dynamical friction switched off, the same transformation leads to a forced harmonic oscillator,
\begin{equation}
\ddot{u}+\kappa^2 u=F(t),
\label{eq:fho}
\end{equation}
where $F(t)$ is a driving term set by the oscillation frequency of the excited mode. For the coefficient $c_{1,1,-1}(t)$, which we found to be most strongly associated with stone skipping, a Fourier analysis of the simulation data indicates a prominent splitting into two comparably strong frequency peaks, suggestive of an effective level splitting. The two dominant frequencies in Figure~\ref{fig:14} are
\begin{equation}
\begin{aligned}
f_1=0.000999983 \quad \mathrm{Myr}^{-1},\\f_2=-0.000233329 \quad \mathrm{Myr}^{-1},
\end{aligned}
\end{equation}
with a separation
\begin{equation}
\Delta f\equiv |f_2-f_1|=1.23\times 10^{-3} \quad \mathrm{Myr}^{-1}.
\end{equation}
For the ordering shown in Figure~\ref{fig:14}, the signed difference is $f_2-f_1=-1.233312\times 10^{-3}\,\mathrm{Myr}^{-1}$.

Empirically, $\Delta f$ coincides with the strongest nonzero frequency in the modulation of $|c_{1,1,-1}(t)|^2$. The corresponding beat angular frequency of the complex coefficient is
\begin{equation}
\omega_{\rm dip}\equiv \frac{|\omega_2-\omega_1|}{2}=\pi\Delta f.
\label{eq:wdip}
\end{equation}
This frequency sets the apparent oscillation frequency of the dipole modes in the time domain. These observations motivate the approximation
\begin{equation}
c_{1,1,-1}(t)\approx A\left(e^{i\omega_1 t}+e^{i\omega_2 t}\right)
=2A\cos(\omega_{\rm dip} t)\,e^{i\bar{\omega} t},
\label{eq:split}
\end{equation}
with carrier frequency
\begin{equation}
\bar{\omega}=\frac{\omega_1+\omega_2}{2}.
\end{equation}

Building on this structure, one can derive an explicit expression for the forcing term $F(t)$ (see Appendix~\ref{apx:C}),
\begin{equation}
F(t)=A_1\sin(\nu_1 t+\delta_1)+A_2\sin(\nu_2 t+\delta_2),
\label{eq:Ft}
\end{equation}
where $\nu_{1,2}$ are the effective forcing frequencies after the dipole time dependence is projected along the nearly circular orbit, and $\delta_{1,2}$ are constant phases. In the absence of dynamical friction, the corresponding resonance condition is
\begin{equation}
\nu_{1,2}^2=\kappa^2=\Phi_0''(r_0)+3\Omega_0^2.
\label{eq:reson0}
\end{equation}

For the most general case in which both dynamical friction and the excited mode are present, the linearized dynamics can be written as a non-standard higher-order differential equation of the form
\begin{equation}
\dddot{\xi}+(\gamma_\phi+\gamma_r)\ddot{\xi}
+\left(\kappa^2+\gamma_\phi\gamma_r\right)\dot{\xi}
+\gamma_\phi\kappa^2\xi
=G(t),
\label{eq:highorder}
\end{equation}
where $G(t)$ is defined in Appendix~\ref{apx:C} and is more complicated than the forcing term in Eq.~\eqref{eq:Ft}. While such a system can still exhibit resonant behavior, a complete  analysis is beyond the scope of the present work. Instead, we consider an instructive limit: we retain radial damping ($\gamma_r\neq 0$) and the mode excitation ($\epsilon\neq 0$), but switch off azimuthal damping by setting $\gamma_\phi=0$. In this limit we have a standard forced, damped harmonic oscillator,
\begin{equation}
\ddot{u}+2\Gamma \dot{u}+\kappa^2 u=F(t),
\label{eq:fdho}
\end{equation}
with $\Gamma=\gamma_r/2$.

For the canonical monochromatic forcing problem
\begin{equation}
\ddot{u}+2\Gamma \dot{u}+\kappa^2 u=F(\omega),
\label{eq:mono}
\end{equation}
the non-decaying steady-state response takes the form
\begin{equation}
\xi(\omega)=
\frac{(a\omega+b)\sin\!\left(\omega t+\delta-\frac{\pi}{2}\right)}
{\omega\sqrt{\left(\kappa^2-\omega^2\right)^2+\left(2\Gamma\omega\right)^2}},
\label{eq:resp}
\end{equation}
where $\delta$ is the phase of the complex amplitude. If $0<\omega\ll-\frac{2\Omega_0\mathrm{\Phi}_{1m}\left(r_0\right)}{r_0\mathrm{\Phi}_{1m}^\prime\left(r_0\right)}$ (see Appendix~\ref{apx:C}), one obtains the resonance condition
\begin{equation}
\omega^2=\frac{4\kappa^2}{3}\left[
\frac{1}{2}-\left(\frac{\Gamma}{\kappa}\right)^2
+\sqrt{\left(\left(\frac{\Gamma}{\kappa}\right)^2-\frac{1}{2}\right)^2-\frac{3}{16}}
\right]
\label{eq:reson2}
\end{equation}
where
\begin{equation}
    \frac{\Gamma}{\kappa}<\frac{1}{2\sqrt{2+\sqrt{3}}}.
\end{equation}
 
In the weak-damping regime this simplifies to
\begin{equation}
\omega^2\simeq \kappa^2-4\Gamma^2,
\qquad
\frac{\Gamma}{\kappa}\ll 1.
\label{eq:reson_weak}
\end{equation}

If $\omega\gg-\frac{2\Omega_0\mathrm{\Phi}_{1m}\left(r_0\right)}{r_0\mathrm{\Phi}_{1m}^\prime\left(r_0\right)}>0$, the resonance condition becomes
\begin{equation}
    \omega=\sqrt{\kappa^2-2\Gamma^2},\qquad  \frac{\Gamma}{\kappa}< \frac{1}{\sqrt{2}}
    \label{eq:reson3}
\end{equation}
Because the forcing term $F(t)$ in Eq.~\eqref{eq:Ft} contains two effective frequencies, $\nu_{1,2}$, the system can resonate at either (or both) of these components, leading to a substantial growth of the deviation from circular motion. Within this framework, stone skipping can be interpreted as an orbital resonance driven by the dipole modes. The model also helps explain the   results: although radial modes can be strongly excited, their dominant frequencies typically lie outside the range of $\omega$ that satisfies the resonance condition~\eqref{eq:reson2} for the configurations studied here.

\section{Conclusion}
\label{sec:VI}

We have investigated the orbital dynamics of black holes immersed in ULDM solitons and identified a mechanism for stone skipping~\cite{wang2022}, in which the orbital radius undergoes long-lived quasi-periodic rebounds after an initial inspiral. Using an eigenmode decomposition and perturbation-theory framework~\cite{zagorac2022}, we find that this behaviour is tightly correlated with a dipole-like excitation of the soliton. Radial modes and higher multipoles  do not reproduce the large repeated rebounds, while adding the dipole sector does. The fully coupled seeded-binary  further shows that the rebound can occur without inserting the empirical drag term used in the reduced diagnostic model.  

We developed a semi-analytic model that captures the essential physics in a restricted near-circular regime. Mapping the dynamics near a reference circular orbit onto a forced oscillator shows that the time-dependent dipole modes yield a periodic driving term while damping supplies dissipation. Within this interpretive framework, stone skipping arises when the forcing contains frequency support close to the orbital or epicyclic frequency and the system enters a resonance window in which energy transfer from the excited soliton to the orbit overcomes the secular damping. The model is not a calibrated replacement for the fully coupled Schr\"odinger-Poisson calculation, but it explains why the dipole phase and frequency content are dynamically important.

Our results show that ULDM orbital decay cannot always be reduced to a quasi-static Chandrasekhar-type drag with a slowly varying Coulomb logarithm~\cite{chandrasekhar1943,chandrasekhar1943a}. The coherent soliton response can generate time-dependent forces and feedback effects that are outside such a local prescription~\cite{lancaster2020, hui2017}. Interestingly, this dipole-driven mechanism has an analogue in classical stellar dynamics, where the $\ell=1$ response has previously been recognized as a source of weakly damped oscillations, or seiche modes, that can significantly affect orbital decay~\cite{weinberg1989, weinberg1994, weinberg2023}. Our findings extend this picture to the wave-mechanical context of ULDM, where the coherent condensate can make such resonant responses especially clear in idealized simulations.

This work has particular relevance to SMBH binaries in galactic nuclei where ULDM-enhanced drag might alleviate the final parsec problem and modify gravitational-wave signals~\cite{koo2024, boey2025,Tiruvaskar:2025hxl}. Our findings show that backreaction-driven resonances can qualitatively prolong orbital evolution even when the mean trend is dissipative. However, full dynamical simulations suggest that stone skipping is not effective in scenarios where the black hole mass is a significant fraction ($\gtrsim 10\%)$ of the soliton mass~\cite{wang2022}. Consequently, this mechanism is most relevant to smaller halos, which tend to have proportionately smaller SMBH:soliton mass ratios~\cite{NANOGrav:2023gor}. A pulsar-timing stochastic background is likely to be dominated by the largest halos, so any observational consequences of stone skipping for merger rates and associated gravitational-wave production are likely more relevant to LISA~\cite{duque2024}. Conversely, equal-mass binaries do not stone skip in the idealized unperturbed-soliton initial conditions studied here, although an externally seeded dipole can change that conclusion.

Several directions follow naturally. A first priority is to embed the soliton in a more realistic galactic environment by incorporating an outer halo, potentially using efficient wave-halo construction methods~\cite{yavetz2022}, and assessing the impact of ambient ULDM granule fluctuations, relaxation processes, and external tidal perturbations~\cite{baror2019, hui2017, widmark2024} on any dipole excitation of the soliton. Second, the resonance model should be generalized to eccentric or inclined orbits and multi-body configurations. Third, because unequal-mass binaries or mergers could excite a dipole term, it will be valuable to delineate the parameter space in which soliton-driven resonance works against a drag-driven inspiral. Fourth, this work can be generalized to mixed cold-plus-ultralight dark matter~\cite{Schwabe:2020eac} and multi-component ULDM models~\cite{Tellez-Tovar:2021mge,Guo:2020tla,Amin:2022pzv,Gosenca:2023yjc}. Finally, the original excitation of the dipole and other modes by orbiting black holes has not been described within perturbation theory; a full treatment of this excitation process would be useful.

\begin{acknowledgments}
AZ is grateful to Brian Schmidt for his mentorship and guidance, and for facilitating the exchange visit to the University of Auckland where this work was initiated. We thank the Department of Physics at the University of Auckland for its hospitality. We also acknowledge the use of computing resources provided by the Australian National University. RE acknowledges support from the Marsden
Fund Council grant MFP-UOA2131 from New Zealand
Government funding, managed by the Royal Society
Te Ap\=arangi and the use of New
Zealand eScience Infrastructure (NeSI) high-performance
computing facilities. YW acknowledges the computing time granted by the Resource Allocation Board and provided on the supercomputer Emmy at NHR-Nord@Göttingen as part of the NHR infrastructure. Some simulations were conducted with computing resources under the project nip00084.
Research at Perimeter Institute is supported in part by the Government of Canada through the Department of Innovation, Science and Economic Development Canada and by the Province of Ontario through the Ministry of Colleges and Universities. At McGill University LZ is supported by the Trottier Space Institute Fellowship.
\end{acknowledgments}

\appendix
\section{Eigenstates of the ULDM soliton}
\label{apx:A}
We review the numerical solution of the eigenfunction equations of the   time-independent Schr\"odinger equation~\eqref{eq:tise} for soliton configurations. In this Appendix we set $m=\hbar=G=c=1$ so Eq.~\eqref{eq:tise} becomes
\begin{equation}
    \begin{aligned}
        \widehat{H}_{0}|n, \ell, m\rangle & =\left(-\frac{1}{2} \nabla^{2}+\Phi_0\right)|n, \ell, m\rangle \\
& =E_{nl}|n, \ell, m\rangle.
    \end{aligned}
    \label{eq:A1}
\end{equation}
Writing the position representation of the ket $\left|n\ell m\right\rangle$ in spherical coordinates as  $\psi_{n\ell m}(\textbf{r})$ 
\begin{equation}
    \begin{aligned}
|n,\ell, m\rangle=\int_{\mathbb{R}^{3}} d^{3} \boldsymbol{r} \psi_{n \ell m}(\boldsymbol{r})|\boldsymbol{r}\rangle.
    \end{aligned}
    \label{eq:A2}
\end{equation}
Taking the $\left\langle \textbf{r}\right|$ derivative of equation \eqref{eq:A1} yields:
\begin{equation}
    \nabla^2\psi_{n\ell m}(\textbf{r})=2\left(\Phi_0-E_{n\ell}\right)\psi_{n\ell m}(\textbf{r}) \, .
    \label{eq:A3}
\end{equation}

Following the standard approach for hydrogen atom wavefunctions, we work in spherical coordinates and employ separation of variables, so that
\begin{equation}
    \begin{aligned}
        &{\left[\frac{\partial_{r}\left(r^{2} \partial_{r}\right)}{r^{2}}\right.}  \left.+\frac{\partial_{\theta}\left(\sin \theta \partial_{\theta}\right)}{r^{2} \sin \theta}+\frac{\partial_{\phi}^{2}}{r^{2} \sin ^{2} \theta}\right] \psi_{n \ell m}(r, \theta, \phi) \\
& =2\left(\Phi_0-E_{n \ell}\right) \psi_{n \ell m}(r, \theta, \phi) ,
    \end{aligned}
    \label{eq:A4}
\end{equation}
and
\begin{equation}
    \psi_{n \ell m}(r, \theta, \phi)=f_{n \ell}(r) \Theta_{\ell m}(\theta)  \widetilde{\Phi}_{m}(\phi).
    \label{eq:A5}
\end{equation}
 The angular portion yields the condition:
\begin{equation}
    \frac{\partial_{\theta}\left(\sin \theta \partial_{\theta} \Theta_{\ell m}(\theta)\right)}{\sin \theta  \widetilde{\Phi}_{m}(\phi)}+\frac{\partial_{\phi}^{2}  \widetilde{\Phi}_{m}(\phi)}{\sin ^{2} \theta \Theta_{\ell m}(\theta)}=-\ell(\ell+1)=C,
    \label{eq:A6}
\end{equation}
where $C$ is a constant.
The radial equation becomes:
\begin{equation}
    \frac{\partial_{r}\left(r^{2} \partial_{r} f_{n \ell}(r)\right)}{r^{2} f_{n \ell}(r)}-\frac{\ell(\ell+1)}{r^{2}}=2\left(\Phi_0-E_{n \ell}\right) .
    \label{eq:A7}
\end{equation}
The solution to equation \eqref{eq:A6} consists of the familiar spherical harmonics:
\begin{equation}
    Y_{\ell m}(\theta, \phi)=\sqrt{\frac{(2 l+1)}{4 \pi} \frac{(\ell-m)!}{(\ell+m)!}}(-1)^{m} P_{\ell m}(\cos \theta) e^{i m \phi}.
    \label{eq:A8}
\end{equation}
 For the radial component, we solve equation \eqref{eq:A7} using a substitution:
\begin{equation}
    u_{n \ell}\left(r\right)=rf_{n \ell}\left(r\right).
    \label{eq:A9}
\end{equation}
This allows us to rewrite:
\begin{equation}
    \frac{\partial_{r}\left(r^{2} \partial_{r} f_{n \ell}(r)\right)}{r^{2}}=\frac{1}{r} \frac{\partial^{2} u_{n \ell}}{\partial r^{2}}.
    \label{eq:A10}
\end{equation}
Equation \eqref{eq:A7} then transforms to:
\begin{equation}
    \frac{\partial^{2} u_{n \ell}}{\partial r^{2}}-\frac{l(l+1)}{r^{2}} u_{n \ell}(r)=2\left(\Phi_0-E_{n \ell}\right) u_{n \ell}(r)
    \label{eq:A11}
\end{equation}
Alternatively, defining $\chi_{\ell}\left(r\right)$ for a more concise form:
\begin{equation}
    \begin{aligned}
        &\frac{1}{2}\left(\chi_{l}(r)-\frac{\partial^{2}}{\partial r^{2}}\right) u_{n \ell}(r)=E_{n \ell} u_{n \ell}(r), \\
&\chi_{\ell}(r)=\ell(\ell+1)r^{-2}+2\Phi_0.
    \end{aligned}
    \label{eq:A12}
\end{equation}
We discretize equation \eqref{eq:A12}, transforming it into a matrix eigenvalue problem. Note that $u_{n \ell}\left(r\right)$ represents a one-dimensional distribution. We impose boundary conditions on the interval $\left[0,r_{max}\right]$, with uniform spacing $\Delta r=r_{i+1}-r_i$, where $i\in\left\{0,1,\cdots N,N+1\right\}$ corresponds to sampling at $N+2$ points, or
\begin{equation}
     r_{i}=i \Delta r, \quad r_{i}+k \Delta r=r_{i+k} , \quad
\Delta r= \frac{r_{\max }}{N+1} .
      \label{eq:A13}
\end{equation}

We can express $u_{n \ell}\left(r\right)$ in vector form:
\begin{equation}
    u_{n \ell}=\left[\begin{array}{llllll}
0 & u_{n \ell}\left(r_{1}\right) & u_{n \ell}\left(r_{2}\right) & \cdots & u_{n \ell}\left(r_{N}\right) & 0 
\end{array}\right]^{T}
\label{eq:A14}
\end{equation}
The boundary conditions are then
\begin{equation}
    u_{n \ell}\left(r_0\right)=r_0f_{n \ell}\left(r_0\right)=0\times f_{n \ell}\left(0\right)=0,
    \label{eq:A15}
\end{equation}
\begin{equation}
    u_{n \ell}\left(r_{N+1}\right)=r_{N+1}f_{n \ell}\left(r_{N+1}\right)=r_{N+1}\times0=0.
    \label{eq:A16}
\end{equation}
Here, $u_{n \ell}\left(r_0\right)$ vanishes by definition of $u_{n \ell}$, while $u_{n \ell}\left(r_{N+1}\right)$ is zero to ensure that the wavefunction $f_{n \ell}$ vanishes at the box edge. For the second derivative  in equation \eqref{eq:A12}, we discretize as follows:
\begin{equation}
    \begin{aligned}
        \frac{d^{2} u}{d r^{2}}:=&\lim _{\Delta r \rightarrow 0} \frac{u(r)-2 u(r+\Delta r)+u(r+2 \Delta r)}{\Delta r^{2}} \\
& \xrightarrow{r \rightarrow r_{i-1}} \frac{u\left(r_{i-1}\right)-2 u\left(r_{i}\right)+u\left(r_{i+1}\right)}{\Delta r^{2}}
    \end{aligned}
    \label{eq:A17}
\end{equation}
This converts equation \eqref{eq:A6} into a coupled set of $N+2$ linear equations. 
\begin{widetext}
\begin{equation}
\begin{aligned}
        \left(\chi_{\ell}\left(r_{i}\right) u_{n \ell}\left(r_{i}\right)-\frac{u\left(r_{i-1}\right)-2 u\left(r_{i}\right)+u\left(r_{i+1}\right)}{\Delta r^{2}}\right)=2E_{n \ell} u_{n \ell}\left(r_{i}\right),
\end{aligned}
    \label{eq:A18}
\end{equation}
where $i \in\{0,1,2 \cdots, N, N+1\}$. The  equations for $r_0$ and $r_{N+1}$ and fixed by the boundary conditions \eqref{eq:A15} and \eqref{eq:A16}, leaving $N$ independent equations 
\begin{equation}
\adjustbox{max width=\textwidth}{$
0.5\left(\left[\begin{matrix}\chi_{\ell}\left(r_1\right)&0&0&\cdots&0\\0&\chi_{\ell}\left(r_2\right)&0&\cdots&0\\\vdots&\ddots&\ddots&\ddots&0\\\vdots&\ddots&\ddots&\chi_{\ell}\left(r_{N-1}\right)&0\\0&\cdots&\cdots&0&\chi_{\ell}\left(r_N\right)\\\end{matrix}\right]-\frac{1}{\Delta r^2}\left[\begin{matrix}-2&1&0&\cdots&0\\1&-2&1&\cdots&0\\\vdots&\ddots&\ddots&\ddots&0\\\vdots&\ddots&1&-2&1\\0&\cdots&\cdots&1&-2\\\end{matrix}\right]\right)\left[\begin{array}{c}u_{n \ell}\left(r_{1}\right) \\ u_{n \ell}\left(r_{2}\right) \\ u_{n \ell}\left(r_{3}\right) \\ \vdots \\ u_{n \ell}\left(r_{N-1}\right) \\ u_{n \ell}\left(r_{N}\right)\end{array}\right]=E_{n\ell}\left[\begin{array}{c}u_{n \ell}\left(r_{1}\right) \\ u_{n \ell}\left(r_{2}\right) \\ u_{n \ell}\left(r_{3}\right) \\ \vdots \\ u_{n \ell}\left(r_{N-1}\right) \\ u_{n \ell}\left(r_{N}\right)\end{array}\right]
$}.
\label{eq:A19}
\end{equation}
\end{widetext}
This is now a standard matrix eigenvalue problem, with eigenvectors $u_{n \ell}$ and eigenvalues $E_{nl}$ 
\begin{equation}
    A_l\,u_{n \ell}=E_{n\ell}\,u_{n \ell},
    \label{eq:A20}
\end{equation}
where $A_l$ is a real, symmetric $N\times N$ matrix. We order the eigenvalues $E_{0\ell},\ldots,E_{N-1,\ell}$ from smallest to largest. The eigenvectors $u_{n \ell}$ are related to the original radial eigenfunctions via
$f_{n \ell}(r)=u_{n \ell}(r)/r$, so 
\begin{equation}
    \psi_{n \ell m}(\boldsymbol{r})=f_{n \ell}(r) Y_{\ell m}(\theta, \phi) .
    \label{eq:A21}
\end{equation}
 A potential source of confusion is to conflate the continuum limit of the \emph{discretization} with the \emph{free-space} limit of the underlying eigenvalue problem. Refining the grid ($N\to\infty$ at fixed $r_{\max}$) does not introduce a missing continuous sector: it simply yields a more accurate representation of the same eigenvalue problem posed on a finite interval with the boundary conditions adopted here. For fixed $\ell$ and fixed $r_{\max}$, Eq.~\eqref{eq:radial} is a regular self-adjoint Sturm--Liouville problem on $[0,r_{\max}]$, and therefore has a purely discrete spectrum with a complete orthonormal set of radial eigenfunctions $u_{n\ell}(r)$; correspondingly, the resolution of the identity in Eq.~\eqref{eq:compt} is purely discrete in this setting. A continuous contribution (schematically $\int d\lambda\,|\lambda\rangle\langle\lambda|$) becomes relevant only in the free-space limit $r_{\max}\to\infty$, where the spectral structure changes.
 
\section{Orbital resonance of black holes}
\label{apx:C}

\subsection{Lagrangian for the Black Hole}
 We treat the black hole as a classical point particle of mass $M$ moving in two-dimensional polar coordinates $(r,\phi)$ and confine its motion to the equatorial plane ($\theta=\pi/2$) throughout.  In polar coordinates, the position is 
\begin{equation}
    \mathbit{X}\left(t\right)=r\left(t\right){\hat{e}}_r\left(\phi\left(t\right)\right).
    \label{eq:C7}
\end{equation}
and the kinetic energy is then the usual sum of radial and azimuthal contributions,
\begin{equation}
    T=\frac{1}{2}M_{BH}\left({\dot{r}}^2+r^2{\dot{\phi}}^2\right).
    \label{eq:C8}
\end{equation}
The potential energy is decomposed into a spherically symmetric ULDM ground-state potential plus a perturbation associated with a given excited mode:
\begin{equation}
    V=M\mathrm{\Phi}_U\left(r,\phi,t\right)=M\left[\mathrm{\Phi}_0\left(r\right)+\delta\mathrm{\Phi}_U\left(\mathbit{r},t\right)\right].
    \label{eq:C9}
\end{equation}
Critically, the perturbation also modifies the gravitational potential. Given equation~\ref{eq:A21} we can 
write a time-dependent wavefunction as a linear superposition,
\begin{equation}
    \Psi(r,\theta,\phi,t)=\sum_{n\ell m}c_{n\ell m}(t)
    \psi_{n\ell m}(r,\theta,\phi),
    \label{eq:C11}
\end{equation}
where the coefficients $c_{n\ell m}(t)$ encode the time dependence. The ground-state soliton corresponds to the mode with $n=\ell=m=0$,
\begin{equation}
    \Psi_0(r)=f_{00}(r)\,Y_{00},
    \label{eq:C12}
\end{equation}
and add a single mode with amplitude $\epsilon$:
\begin{equation}
\begin{aligned}
&\mathrm{\Psi}\left(\mathbit{r},t\right)=\mathrm{\Psi}_0+\epsilon\cdot\delta\mathrm{\Psi}^{nlm}\left(\mathbit{r},t\right),\\ &\delta\mathrm{\Psi}^{nlm}\left(\mathbit{r},t\right)=c_{nlm}\left(t\right)f_{nl}\left(r\right)Y_{lm}\left(\theta,\phi\right).
    \label{eq:C13}
\end{aligned}
\end{equation}

We excite a single dipole mode with $\ell=1$:
\begin{equation}
\delta\mathrm{\Psi}^{n1m}\left(\mathbit{r},t\right)=c_{n1m}\left(t\right)f_{n1}\left(r\right)Y_{1m}\left(\theta,\phi\right).
    \label{eq:C15}
\end{equation}
In principle one could examine the coupling between the soliton eigenmodes and an orbiting mass but we will inject the empirical time dependence of $c_{n1m}\left(t\right)$ we found from numerical simulations (see Figure \ref{fig:14}), or
\begin{equation}
\begin{aligned}
&c_{n1m}\left(t\right)\approx C\cos{\left(\omega_{\rm dip}t\right)}e^{i\bar{\omega}t},\\ &\bar{\omega}=\frac{\omega_1+\omega_2}{2},\\ &\omega_{\rm dip}=\frac{|\omega_2-\omega_1|}{2},
\end{aligned}
\label{eq:C16}
\end{equation}
where $C$ is a complex constant and $\omega_{\rm dip}$ is the positive beat angular frequency of the mode coefficient. Consequently, we have
\begin{equation}
    \delta\Psi\approx C\epsilon\cos{\left(\omega_{\rm dip}t\right)}e^{i\bar{\omega}t}f_{n1}\left(r\right)Y_{1m}\left(\theta,\phi\right)
    \label{eq:C17}
\end{equation}
 
The ULDM density is given by
\begin{equation}
\rho\left(\mathbit{r},t\right)=\left|\mathrm{\Psi}\left(r,t\right)\right|^2.
    \label{eq:C18}
\end{equation}
Substituting $\mathrm{\Psi}\left(\mathbit{r},t\right)=\mathrm{\Psi}_0+\delta\Psi$ and linearizing yields
\begin{equation}
|\mathrm{\Psi}_0+\delta\Psi|^{2}=|\mathrm{\Psi}_0|^{2}+\mathrm{\Psi}_0\delta\Psi^{\ast}+\mathrm{\Psi}_0^{\ast}\delta\Psi+\mathcal{O}(|\delta\Psi|^{2}).
\label{eq:C19}
\end{equation}
Thus, the density perturbation can be written as
\begin{equation}  \delta\rho=\mathrm{\Psi}_0^\ast\delta\Psi+\mathrm{\Psi}_0\delta\mathrm{\Psi}^\ast=2\mathfrak{R}\left[\mathrm{\Psi}_0^\ast\delta\Psi\right].
    \label{eq:C20}
\end{equation}
Substituting \eqref{eq:C17}, we find:
\begin{equation}
\begin{aligned}
\delta\rho&=2\mathfrak{R}\left[\mathrm{\Psi}_0^\ast C\cos{\left(\omega_{\rm dip}t\right)}\epsilon e^{i\bar{\omega}t}f_{n1}\left(r\right)Y_{1m}\left(\theta,\phi\right)\right]\\&=2\left|C\right|\left|\epsilon\right|\mathrm{\Psi}_0f_{n1}\left(r\right)\cos{\left(\omega_{\rm dip}t\right)}\mathfrak{R}\left[e^{i\alpha}e^{i\bar{\omega}t}Y_{1m}\left(\theta,\phi\right)\right],
\end{aligned}
\label{eq:C21}
\end{equation}
where the ground-state wavefunction $\mathrm{\Psi}_0$ and the radial part $f_{n1}(r)$ of the excited mode are both taken to be real and $\alpha$ is defined as the phase of the product $C\epsilon$. Since we restrict to equatorial-plane motion ($\theta=\pi/2$) $Y_{1m}(\theta,\phi)$ reduces to
\begin{equation}
    Y_{1m}\!\left(\frac{\pi}{2},\phi\right)=
    \begin{cases}
      \displaystyle\sqrt{\frac{3}{8\pi}}\,e^{-\mathrm{i}\phi},& m=-1,\\[4pt]
      0,& m=0,\\[4pt]
      \displaystyle\sqrt{\frac{3}{8\pi}}\,e^{+\mathrm{i}\phi},& m=+1,
    \end{cases}
    \label{eq:C22}
\end{equation}
and the density perturbation becomes 
\begin{equation}
\begin{aligned}
\delta\rho&=\mathcal{D}(r,t)
    \begin{cases}
      \sqrt{\frac{3}{8\pi}}\mathfrak{R}\left\{e^{i\left(\alpha-\phi+\bar{\omega}t\right)}\right\},& m=-1,\\[4pt]
      0,& m=0,\\[4pt]
      \sqrt{\frac{3}{8\pi}}\mathfrak{R}\left\{e^{i\left(\alpha+\phi+\bar{\omega}t\right)}\right\},& m=+1,
    \end{cases}\\
\mathcal{D}(r,t)&=2\left|C\right|\left|\epsilon\right|\mathrm{\Psi}_0f_{n1}\left(r\right)\cos{\left(\omega_{\rm dip}t\right)} .
\end{aligned}
    \label{eq:C23}
\end{equation}
A corollary of this choice is that the $m=0$ dipole mode is not relevant to stone skipping. Given rotational symmetry, we absorb $\alpha\pm\phi$ into $\alpha$ by a redefinition of the azimuthal origin and write  
\begin{equation}
\delta\rho_\pm=2\left|C\right|\left|\epsilon\right|f_{00}\left(r\right)f_{n1}\left(r\right)\cos{\left(\omega_{\rm dip}t\right)}\cos{\left(\phi\pm\bar{\omega}t\right)},
    \label{eq:C24}
\end{equation}
where the subscript plus and minus signs correspond respectively to $m=+1$ and $m=-1$. 
 
The next step is to use the Poisson equation 
\begin{equation}
    \nabla^{2}\delta\Phi=4\pi G\,\delta\rho,
    \label{eq:C25}
\end{equation}
to convert $\delta\rho_{\pm}$
into a gravitational potential perturbation $\delta\Phi$. Again expanding  in spherical harmonics  
\begin{equation}
    \delta\rho=\sum_{\ell,m}\rho_{\ell m}Y_{\ell m}, \quad
       \delta\Phi=\sum_{\ell,m}\Phi_{\ell m}Y_{\ell m}
\label{eq:C26}
\end{equation}
and substituting into Eq.~\eqref{eq:C24}  orthogonality shows each $(\ell,m)$ mode satisfies its own radial differential equation,
\begin{equation}
    \frac{1}{r^{2}}\frac{\mathrm{d}}{\mathrm{d}r}\!\left(r^{2}\frac{\mathrm{d}\Phi_{\ell m}}{\mathrm{d}r}\right)
    -\frac{\ell(\ell+1)}{r^{2}}\Phi_{\ell m}=4\pi G\,\rho_{\ell m}.
    \label{eq:C27}
\end{equation}

Our focus is $(\ell,m)=(1,m)$ and  it is  convenient to write
\begin{equation}
\delta\Phi=\epsilon_\Phi\mathrm{\Phi}_{1m}\cos{\left(\omega_{\rm dip}t\right)}\cos{\left(\phi\pm\bar{\omega}t\right)},
    \label{eq:C28}
\end{equation}
where $\epsilon_\Phi$ absorbs all constant prefactors; for brevity we  write $\epsilon$  in what follows. 

The full Lagrangian for a black hole moving in a ULDM soliton with an $\ell=1$ dipole perturbation is
\begin{equation}
\begin{aligned}
L&=\frac{1}{2}M_{BH}\left({\dot{r}}^2+r^2{\dot{\phi}}^2\right)\\&-M_{BH}\left[\mathrm{\Phi}_0\left(r\right)+\epsilon\mathrm{\Phi}_{1m}\left(r\right)\cos{\left(\omega_{\rm dip}t\right)}\cos{\left(\phi-\bar{\omega}t\right)}\right].
\end{aligned}
\label{eq:C29}
\end{equation}
In Eq. \eqref{eq:C29} we choose the combination $\phi-\bar{\omega}t$, corresponding to the $m=-1$ mode.

Given the presence of non-conservative drag forces  $Q_i$, 
\begin{equation}
    \frac{\mathrm{d}}{\mathrm{d}t}\frac{\partial\mathcal{L}}{\partial \dot{q}_i} - \frac{\partial \mathcal{L}}{\partial q_i} = Q_i .
    \label{eq:C30}
\end{equation}
In planar polar coordinates   $\mathbf{f}=f_r\,\hat{\mathbf{e}}_r + f_\phi\,\hat{\mathbf{e}}_\phi$, so that
$Q_r=F_r$ and $Q_\phi = r F_\phi$. The   equations of motion are thus
\begin{widetext}
    \begin{align}
      &\ddot{r}-r{\dot{\phi}}^2+\frac{d\mathrm{\Phi}_0\left(r\right)}{dr}+\epsilon\frac{d\mathrm{\Phi}_{1m}\left(r\right)}{dr}\cos{\left(\omega_{\rm dip}t\right)}\cos{\left(\phi-\bar{\omega}t\right)}=\frac{Q_r}{M_{BH}},  \label{eq:C31a}\\ 
      &\frac{d}{dt}\left(r^2\dot{\phi}\right)-\epsilon\mathrm{\Phi}_{1m}\left(r\right)\cos{\left(\omega_{\rm dip}t\right)}\sin{\left(\phi-\bar{\omega}t\right)}=\frac{Q_\phi}{M_{BH}}.  \label{eq:C31b}
    \end{align}
\end{widetext}
For tractability we assume that the frictional force is  proportional to the velocity, 
\begin{equation}
\begin{aligned}
    f_r=-M_{BH}\gamma_r\dot{r},\\
    f_\phi=-M_{BH}\gamma_\phi r\dot{\phi},
\end{aligned}
\label{eq:C32}
\end{equation}
introduce damping coefficients $\gamma_r$ and $\gamma_\phi$. Taking the above expressions and equations \eqref{eq:C32} and  \eqref{eq:df}, 
\begin{equation}
\gamma_{r,\phi}=A\frac{4\pi\rho\left(r,t\right)G^2M_{BH}m^2r^2}{3\hbar^2\sqrt{{\dot{r}}^2+r^2{\dot{\phi}}^2}}.
\label{eq:C33}
\end{equation}
 the generalized forces become 
\begin{equation}
\begin{aligned}
Q_r&=-M_{BH}\gamma_r\dot{r},\\
Q_\phi&=-M_{BH}\gamma_\phi r^2\dot{\phi}.
\end{aligned}
\label{eq:C34}
\end{equation}

As a starting point it is useful to consider the simplest limit: an unperturbed circular orbit with no excitation and no dynamical friction, {\em i.e.} $\Phi=0$ and $\gamma_r=\gamma_\phi=0$, which gives  
\begin{equation}
        \ddot{r}-r{\dot{\phi}}^2+\frac{d\mathrm{\Phi}_0\left(r\right)}{dr}=0, \quad
      \frac{d}{dt}\left(r^2\dot{\phi}\right)=0.
    \label{eq:C35}
\end{equation}
To restrict to circular orbits we set $r(t)=r_0$ and
$\dot{r}=0$, and write $\phi(t)=\Omega_0 t + \phi_0$.  The radial part of Eq.~\eqref{eq:C35} then yields the circular-orbit
potential--frequency relation
\begin{equation}
    r_0\Omega_0^2=\mathrm{\Phi}_0^\prime\left(r_0\right).
    \label{eq:C36}
\end{equation}
The azimuthal equation reproduces the usual conservation of angular momentum and will not be needed. 

\subsection{Linearised quasi-circular expansion }

Once we include dynamical friction the orbits become quasi-circular. We write the perturbation as
\begin{equation}
    r(t) = r_0 + \xi(t),\qquad \phi(t) = \Omega_0 t + \eta(t),
    \label{eq:C37}
\end{equation}
where the perturbations $\xi,\eta$ are first order, or 
    $|\xi| \ll r_0$ and  $|\eta|\ll1$.
Under these assumptions we expand $r\dot{\phi}^2$, the derivative of the spherically symmetric potential $\mathrm{\Phi}_0^\prime\left(r\right)$, and the dipole term $\epsilon\,\Phi_{1m}'(r)\cos(\omega_{\rm dip}t)\cos(\phi-\bar{\omega}t)$ about the circular orbit. Introducing constants
\begin{equation}
    C_r = \Phi_{1m}'(r_0),\qquad C_\phi = \Phi_{1m}(r_0),
    \label{eq:C39}
\end{equation}
to encode the radial profiles of the dipole perturbation, the linearized expansions are
\begin{subequations}
\begin{align}
r\dot{\phi}^2 &\simeq r_0\,\Omega_0^2 + \Omega_0^2\,\xi + 2 r_0\,\Omega_0\,\dot{\eta}, \label{eq:C40a}\\
\Phi_0'(r) &\simeq \Phi_0'(r_0) + \xi\,\Phi_0''(r_0), \label{eq:C40b}\\
\Phi_{1m}'(r) &\simeq \Phi_{1m}'(r_0) + \xi\,\Phi_{1m}''(r_0), \label{eq:C40c}\\
\cos(\phi-\bar\omega t) &\simeq \cos\!\bigl((\Omega_0-\bar\omega)t\bigr) - \eta\,\sin\!\bigl((\Omega_0-\bar\omega)t\bigr). \label{eq:C40d}
\end{align}
\end{subequations}

and, up to $\mathcal{O}(\epsilon^2)$,
\begin{subequations}
\begin{align}
\frac{\mathrm{d}}{\mathrm{d}t}\bigl(r^2\dot{\phi}\bigr)
&\simeq r_0^2\,\ddot{\eta} + 2 r_0\,\Omega_0\,\dot{\xi},
\label{eq:C41a}\\ 
r^2\dot{\phi}
&\simeq - r_0^2\bigl(\Omega_0+\dot{\eta}\bigr)-2\Omega_0 r_0\,\xi,
\label{eq:C41b}\\
\epsilon\,\Phi_{1m}'(r)\cos(\omega_{\rm dip}t)
&\cos\!\bigl(\phi-\bar{\omega}t\bigr) \notag\\
&\simeq \frac{\epsilon C_r}{2}\sum_{i=1}^{2}\cos(\nu_i t+\delta_i),
\label{eq:C41c}\\
\epsilon\,\Phi_{1m}(r)\cos(\omega_{\rm dip}t)
&\sin\!\bigl(\phi-\bar{\omega}t\bigr) \notag\\
&\simeq \frac{\epsilon C_\phi}{2}\sum_{i=1}^{2}\sin(\nu_i t+\delta_i).
\label{eq:C41d}
\end{align}
\end{subequations}

Substituting Eqs.~\eqref{eq:C39}--\eqref{eq:C41d} into the equations of motion and using the circular-orbit identity \eqref{eq:C36}, we obtain the coupled linear system
\begin{equation}
    \begin{cases}
      \ddot{\xi} + \mathcal{A}\,\xi - 2 r_0\,\Omega_0\,\dot{\eta} + \gamma_r\,\dot{\xi} =F_r(t),\\[6pt]
      r_0^2\,\ddot{\eta} + 2 r_0\,\Omega_0\,\dot{\xi} + \gamma_\phi\,r_0^2\,\bigl(\Omega_0+\dot{\eta}\bigr) = F_\phi(t),
    \end{cases}
    \label{eq:C42}
\end{equation}
where for convenience we defined
\begin{equation}
\begin{aligned}
    F_r(t)&\equiv-\frac{\epsilon C_r}{2}\sum_{i=1}^{2}\cos(\nu_i t+\delta_i),\\
    F_\phi(t)&\equiv-\frac{\epsilon C_\phi}{2}\sum_{i=1}^{2}\sin(\nu_i t+\delta_i),\\
    \mathcal{A} &\equiv V''_0(r_0) - \Omega_0^2.
\end{aligned}
    \label{eq:C43}
\end{equation}

Equation~\eqref{eq:C42} is a coupled linear system for $\xi(t)$ and $\eta(t)$.  We first examine the  case with no driving or damping, {\em i.e.} $\epsilon=\gamma_r=\gamma_\phi=0$. In this limit Eq.~\eqref{eq:C42} reduces to the homogeneous system
\begin{equation}
    \begin{cases}
      \ddot{\xi} + \mathcal{A}\,\xi - 2 r_0\,\Omega_0\,\dot{\eta} = 0,\\[6pt]
      r_0^2\,\ddot{\eta} + 2 r_0\,\Omega_0\,\dot{\xi} = 0.
    \end{cases}
    \label{eq:C44}
\end{equation}
Differentiating the first of these equations with respect to time and using the second to eliminate $\ddot{\eta}$ gives
\begin{equation}
    \dddot{\xi} + \mathcal{A}\,\dot{\xi} + 4 \Omega_0^2\,\dot{\xi} = 0.
    \label{eq:C45}
\end{equation}
Defining $u(t)\equiv\dot{\xi}(t)$,
\begin{equation}
    \ddot{u} + \kappa^2\,u = 0,
    \label{eq:C46}
\end{equation}
with intrinsic frequency
\begin{equation}
    \kappa^2 = \mathcal{A} + 4 \Omega_0^2 = \mathrm{\Phi}_0^{\prime\prime}\left(r_0\right) + 3\,\Omega_0^2.
    \label{eq:C47}
\end{equation}
Thus, in the absence of dynamical friction and excitation, the soliton--black-hole system exhibits small oscillations around the circular orbit with natural frequency~$\kappa$.

Next we turn to the case with driving but no damping $(\epsilon\neq0,\gamma_r=\gamma_\phi=0)$.  In this regime Eq.~\eqref{eq:C42} reduces to
\begin{equation}
    \begin{cases}
      \ddot{\xi} + \mathcal{A}\,\xi - 2 r_0\,\Omega_0\,\dot{\eta} =F_r(t),\\[6pt]
      r_0^2\,\ddot{\eta} + 2 r_0\,\Omega_0\,\dot{\xi} = F_\phi(t).
    \end{cases}
    \label{eq:C48}
\end{equation}
Differentiating the first equation yields
\begin{equation}
    \dddot{\xi} + \mathcal{A}\,\dot{\xi} - 2 r_0\,\Omega_0\,\ddot{\eta} ={\dot{F}}_r\left(t\right),
    \label{eq:C49}
\end{equation}
where ${\dot{F}}_r\left(t\right)$ is the time derivative of $F_r(t)$. Solving the second equation of Eq.~\eqref{eq:C48} for $\ddot{\eta}$, we find
\begin{equation}
    \ddot{\eta} = \frac{1}{r_0^2}\bigl[F_\phi(t) - 2 r_0\,\Omega_0\,\dot{\xi}\bigr].
    \label{eq:C50}
\end{equation}
Substituting Eq.~\eqref{eq:C50} into Eq.~\eqref{eq:C49} and using $\kappa^2=\mathcal{A}+4 \Omega_0^2$ gives
\begin{equation}
    \dddot{\xi} + \kappa^2\,\dot{\xi} = \dot{F}_r(t) + \frac{2 \Omega_0}{r_0}\,F_\phi(t).
    \label{eq:C51}
\end{equation}
This gives the forced harmonic oscillator
\begin{equation}
    \ddot{u} + \kappa^2\,u = F(t),
    \label{eq:C52}
\end{equation}
driven by a two-frequency forcing term
\begin{equation}
    \begin{aligned}
        F\left(t\right)&={\dot{F}}_r\left(t\right)+\frac{2\Omega_0}{r_0}F_\phi\left(t\right)\\
        &=F_1\sin(\nu_1 t+\delta_1)+F_2\sin(\nu_2 t+\delta_2),\\
        F_i&=a\nu_i+b,\qquad i=1,2,\\
        a&=\frac{C_r\epsilon}{2},b=-\frac{\Omega_0C_\phi\epsilon}{r_0},
    \end{aligned}
    \label{eq:C53}
\end{equation}
  The resonance condition is simply
\begin{equation}
\begin{aligned}
    \kappa=\nu_i,\qquad F_i\ne0,\qquad i=1,2,
    \label{eq:C54}
\end{aligned}
\end{equation}
and combining Eqs.~\eqref{eq:C47} and~\eqref{eq:C54} yields the forcing frequency required for resonance,
\begin{equation}
    \nu_i^2=\kappa^2=\mathrm{\Phi}_0^{\prime\prime}\left(r_0\right)+3\Omega_0^2.
    \label{eq:C55}
\end{equation}

We now add the damping and in this situation Eq.~\eqref{eq:C42} is our equation of motion. As before, we differentiate the radial equation to obtain
\begin{equation}
    \dddot{\xi} + \mathcal{A}\,\dot{\xi} - 2 r_0\,\Omega_0\,\ddot{\eta} + \gamma_r\,\ddot{\xi} = \dot{F}_r(t),
    \label{eq:C56}
\end{equation}
and use the azimuthal equation to express $\ddot{\eta}$ in terms of $\xi$, $\dot{\xi}$ and the effective azimuthal driving term,
\begin{equation}
\begin{aligned}
    \ddot{\eta} = \frac{1}{r_0^2}\,&\Bigl[F_\phi(t) - 2 r_0\,\Omega_0\,\dot{\xi}  \\&-\gamma_\phi\,r_0^2\,\bigl(\Omega_0 + \dot{\eta}\bigr)-2\gamma_\phi\Omega_0r_0\xi\Bigr].
    \label{eq:C57}
\end{aligned}
\end{equation}
Combining these results leads to
\begin{equation}
\begin{aligned}
    \dot{F}_r(t) + \frac{2 \Omega_0}{r_0}\,F_\phi(t) = &\dddot{\xi} + \kappa^2\,\dot{\xi} + 2 \gamma_\phi\,r_0\,\Omega_0^2 \\
    &+ 2 \gamma_\phi\,r_0\,\Omega_0\,\dot{\eta} +4\Omega_0^2\gamma_\phi\xi+\gamma_r\,\ddot{\xi},
\end{aligned}
\label{eq:C58}
\end{equation}
which still contains $\dot{\eta}$.  To eliminate $\dot{\eta}$ we use the radial equation \eqref{eq:C42} once more to rewrite
\begin{equation}
    2 r_0\,\Omega_0\,\dot{\eta} = \ddot{\xi} + \mathcal{A}\,\xi + \gamma_r\,\dot{\xi} - F_r(t).
    \label{eq:c59}
\end{equation}
Substituting this back yields the third-order equation
\begin{equation}
    \dddot{\xi} + \bigl(\gamma_\phi + \gamma_r\bigr)\,\ddot{\xi} 
    + \bigl(\kappa^2 + \gamma_\phi\,\gamma_r\bigr)\,\dot{\xi}
    +\gamma_\phi\kappa^2\xi = G(t),
    \label{eq:C60}
\end{equation}
where the effective driving term is
\begin{equation}
    G(t) \equiv \dot{F}_r(t) + \gamma_\phi\,F_r(t) + \frac{2 \Omega_0}{r_0}\,F_\phi(t) - 2 \gamma_\phi\,r_0\,\Omega_0^2.
    \label{eq:C61}
\end{equation}

\subsection{Forced damped harmonic oscillator}

Eq.~\eqref{eq:C60} contains a term proportional to $\xi$ in addition to derivatives up to third order, it cannot, in general, be cast into the standard form of a forced damped harmonic oscillator equation for $u=\dot{\xi}$,
\begin{equation}
    \ddot{u} + 2\Gamma\,\dot{u} + \kappa^2\,u = F(t).
    \label{eq:C62}
\end{equation}
On closer inspection, the situation simplifies considerably if we keep $\gamma_r\neq0$ but set $\gamma_\phi=0$.  Physically, $\gamma_\phi=0$ corresponds to switching off azimuthal damping, which mainly controls the secular decay of the orbital angular momentum.  The ``resonant amplification of the radial amplitude'' that we are interested in is essentially a local process, occurring on a timescale short compared with the orbital decay time.  Treating $\Omega_0$ as approximately constant, Eq.~\eqref{eq:C60} reduces to Eq.~\eqref{eq:C62}, where
\begin{equation}
    \Gamma = \frac{1}{2}\,\gamma_r,\qquad
    \kappa^2 =\mathrm{\Phi}_0^{\prime\prime}\left(r_0\right) + 3\,\Omega_0^2,
    \label{eq:C63}
\end{equation}
and the driving term is
\begin{equation}
\begin{aligned}
    F_1(t) =(a\nu_1+b) \sin(\nu_1t+\delta_1),\\
    F_2(t)  =(a\nu_2+b) \sin(\nu_2t+\delta_2).
\end{aligned}
 \label{eq:C64}
\end{equation}
To extract the intrinsic frequency in the damped case, it is convenient to consider the complex form
\begin{equation}
    \ddot{u} + 2\Gamma\,\dot{u} + \kappa^2\,u = \mathrm{Im}\bigl[F(\omega)\,e^{i\omega t}\bigr],
    \label{eq:C65}
\end{equation}
and assume a steady-state particular solution
\begin{equation}
    u(\omega) = \mathrm{Im}\bigl[U(\omega)\,e^{i\omega t}\bigr].
    \label{eq:C66}
\end{equation}
Neglecting the homogeneous solution (which decays on the timescale set by~$\Gamma$), substitution into Eq.~\eqref{eq:C65} yields the complex amplitude
\begin{equation}
    U(\omega) = \frac{F(\omega)}{\kappa^2 - \omega^2 + 2 i \Gamma\,\omega}.
    \label{eq:C67}
\end{equation}
The corresponding physical solution for $u(t)$ is
\begin{equation}
\begin{aligned}
    u(\omega) &= \mathcal{U}(\omega)
    \sin\!\bigl(\omega t + \delta\bigr),
    \\
    \mathcal{U}(\omega)&=\frac{a\omega+b}{\sqrt{(\kappa^2-\omega^2)^2 + (2\Gamma\omega)^2}},\\
    \delta & = \arctan\!\frac{-2\Gamma \omega}{\kappa^2-\omega^2}.
\end{aligned}
\label{eq:C68}
\end{equation}
The resonance condition occurs where the derivation of the non-oscillatory part $\mathcal{U}(\omega)$ for the maximum value occurs. The story is not quite complete, however:  $u=\dot{\xi}$, and we are interested in resonance of the orbital radius itself, not of its derivative.  To address this we define
\begin{equation}
    \xi(\omega) = \mathrm{Im}\bigl[\Xi(\omega)\,e^{i\omega t}\bigr],
    \label{eq:C69}
\end{equation}
which implies
\begin{equation}
    \dot{\xi} = \mathrm{Im}\bigl[\Xi(\omega)\,i\omega\,e^{i\omega t}\bigr] = u,
    \label{eq:C70}
\end{equation}
so that
\begin{equation}
    \Xi(\omega) = -\,\frac{i\,U(\omega)}{\omega}
    = -\,\frac{i\,F(\omega)}{\omega\bigl(\kappa^2 - \omega^2 + 2 i \Gamma\,\omega\bigr)}.
    \label{eq:C71}
\end{equation}
From this it follows that
\begin{equation}
\begin{aligned}
    \xi(\omega) &= \mathcal{X}(\omega)
    \sin\!\bigl(\omega t + \delta-\frac{\pi}{2}\bigr),
    \\
    \mathcal{X}(\omega)&=\frac{a\omega+b}{\omega\sqrt{(\kappa^2-\omega^2)^2 + (2\Gamma\omega)^2}}.
\end{aligned}
\label{eq:C72}
\end{equation}
If there exists an $\omega=\omega_{\max}$ such that
\begin{equation}
\left.\frac{d\mathcal{X}}{d\omega}\right|_{\omega_{\max}}=0,
\qquad
\left.\frac{d^2\mathcal{X}}{d\omega^2}\right|_{\omega_{\max}}<0,
\label{eq:C73}
\end{equation}
then the resonance condition is determined by
\begin{equation}
\omega_R=\omega_{\max}.
\label{eq:A74}
\end{equation}
For the full expression of $\mathcal{X}$, a closed-form solution for $\omega_{\max}$ does not generally exist. However, if the system satisfies
\begin{equation}
\omega \gg -\frac{2\Omega_0\Phi_{1m}(r_0)}{r_0\Phi_{1m}'(r_0)}
=\frac{b}{a}>0,
\label{eq:C75}
\end{equation}
we may approximate
\begin{equation}
\mathcal{X}(\omega)\approx \mathcal{X}_a(\omega)
=\frac{a}{\sqrt{\left(\kappa^2-\omega^2\right)^2+\left(2\Gamma\omega\right)^2}}.
\label{eq:C76}
\end{equation}
This is the standard forced, damped oscillator response, for which resonance occurs at
\begin{equation}
\omega_R=\sqrt{\kappa^2-2\Gamma^2},
\qquad
\frac{\Gamma}{\kappa}<\frac{1}{\sqrt{2}}.
\label{eq:C77}
\end{equation}
In the opposite limit, if
\begin{equation}
0<\omega \ll -\frac{2\Omega_0\Phi_{1m}(r_0)}{r_0\Phi_{1m}'(r_0)},
\label{eq:C78}
\end{equation}
\begin{figure*}
  \centering
  \includegraphics[width=\textwidth]{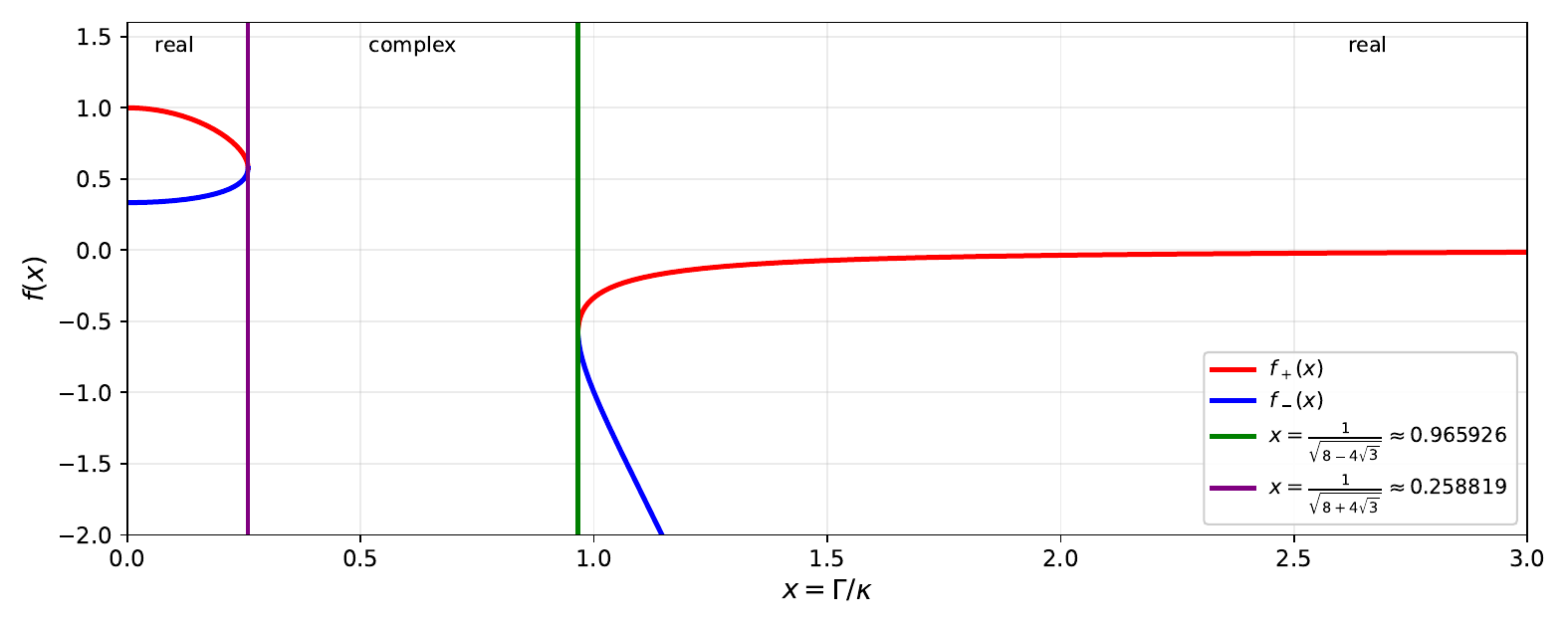}
  \caption{\label{fig:16} Rescaled roots $\mu_\pm/\kappa^2$ of Eq.~\eqref{eq:C85} as functions of $x=\Gamma/\kappa$. The vertical lines mark the boundaries between real and complex roots.}
\end{figure*}
we may instead approximate
\begin{equation}
\mathcal{X}(\omega)\approx \mathcal{X}_b(\omega)
=\frac{b}{\omega\sqrt{\left(\kappa^2-\omega^2\right)^2+\left(2\Gamma\omega\right)^2}}.
\label{eq:C79}
\end{equation}
We define a new denominator function controlling the radial amplitude,
\begin{equation}
    D_1(\omega) \equiv \omega^2\bigl[(\kappa^2-\omega^2)^2 + (2\Gamma\omega)^2\bigr].
    \label{eq:C80}
\end{equation}
Differentiating, we find
\begin{equation}
    \frac{\mathrm{d}D_1(\omega)}{\mathrm{d}\omega} 
     = 6 \omega\Bigl[\omega^4 + \frac{4}{3}\bigl(2\Gamma^2-\kappa^2\bigr)\,\omega^2 + \frac{\kappa^4}{3}\Bigr].
    \label{eq:C81}
\end{equation}

If the quartic polynomial inside the brackets is non\-negative for all $\omega$, then $D_1(\omega)$ attains its minimum at $\omega=0$, and $\xi(\omega)$ would diverge there with no finite maximum at non\-zero $\omega$.  To obtain a genuine resonance at a non\-zero frequency we therefore require that this quartic be negative over some interval in $\omega$, which is equivalent to demanding that the discriminant
\begin{equation}
\begin{aligned}
    \Delta &= \frac{16\bigl(2\Gamma^2-\kappa^2\bigr)^2}{9} - \frac{4\kappa^4}{3}
           \\&= \frac{4\left(\kappa^4-16\kappa^2\Gamma^2+16\Gamma^4\right)}{9}
           \ge 0,
    \label{eq:C82}
\end{aligned}
\end{equation}
or, in terms of the ratio $x=\Gamma/\kappa$,
\begin{equation}
    x^2 \in \bigl[\,0,\,\frac{2-\sqrt{3}}{4}\,\bigr]\;\cup\;\bigl[\,\frac{2+\sqrt{3}}{4},\,\infty\bigr].
    \label{eq:C83}
\end{equation}
Within these ranges, the condition $\mathrm{d}D_1(\omega)/\mathrm{d}\omega=0$ admits three solutions for $\mu=\omega^2$:
\begin{equation}
\begin{aligned}
    \mu_0 &= 0,\\
    \qquad
    \mu_{\pm} &= \frac{2\kappa^2 - 4\Gamma^2 \pm \sqrt{\kappa^4 - 16\kappa^2\Gamma^2 + 16\Gamma^4}}{3}.
    \label{eq:C84}
\end{aligned}
\end{equation}
The two non\-zero solutions can be written in the rescaled form
\begin{equation}
    \frac{\mu_{\pm}}{\kappa^2} 
    = \frac{1}{3}\left[\,2 - 4x^2 \pm \sqrt{1-16x^2+16x^4}\,\right],
    \label{eq:C85}
\end{equation}
as shown in Figure~\ref{fig:16}. The large-damping branch $x^2\ge(2+\sqrt{3})/4\approx0.9659$ does not give the weakly damped resonance relevant to the simulations. On the weak-damping branch $0\le x^2\le(2-\sqrt{3})\approx0.2588$, the larger root $\mu_+$ gives the finite-frequency minimum of $D_1$ after the singular zero-frequency point is excluded. We therefore identify the resonance with
\begin{equation}
    \omega_{d}^2 = \mu_+ = \frac{\kappa^2}{3}\left[\,2 - 4x^2 + \sqrt{1-16x^2+16x^4}\,\right],
    \label{eq:C87}
\end{equation}
which in the small-damping regime $\Gamma/\kappa\ll 1$ reduces to
\begin{equation}
    \omega_{d}^2 \simeq \kappa^2 - 4 \Gamma^2.
    \label{eq:C88}
\end{equation}
Thus, in the low-frequency limit of Eq.~\eqref{eq:C78}, the resonance of the radial amplitude $\xi(\omega)$ occurs when the driving frequency satisfies
\begin{equation}
\begin{aligned}
     \omega_{d}^2 &= \frac{\kappa^2}{3}\left[\,2 - 4x^2 + \sqrt{1-16x^2+16x^4}\,\right],\\
     x&<\frac{1}{2\sqrt{2+\sqrt{3}}}.
\end{aligned}
    \label{eq:C89}
\end{equation}
In the high-frequency limit of Eq.~\eqref{eq:C75}, the corresponding standard forced-oscillator result is
\begin{equation}
    \omega_d^2=\kappa^2-2\Gamma^2,\qquad \frac{\Gamma}{\kappa}<\frac{1}{\sqrt{2}}.
    \label{eq:C90}
\end{equation}

\bibliographystyle{apsrev4-2}
\bibliography{sk}

@PREAMBLE{
 "\providecommand{\noopsort}[1]{}" 
 # "\providecommand{\singleletter}[1]{#1}%" 
}

@ARTICLE{alonso-alvarez2024,
  author = {Alonso-{\'A}lvarez, Gonzalo and Cline, James M. and Dewar, Caitlyn},
  title = {Self-Interacting Dark Matter Solves the {Final Parsec Problem} of {Supermassive Black Hole Mergers}},
  journal = {Phys.\ Rev.\ Lett.},
  volume = {133},
  number = {2},
  pages = {021401},
  year = {2024},
  doi = {10.1103/PhysRevLett.133.021401}
}

@article{NANOGrav:2023gor,
    author = "Agazie, Gabriella and others",
    collaboration = "NANOGrav",
    title = "{The NANOGrav 15 yr Data Set: Evidence for a Gravitational-wave Background}",
    eprint = "2306.16213",
    archivePrefix = "arXiv",
    primaryClass = "astro-ph.HE",
    doi = "10.3847/2041-8213/acdac6",
    journal = "Astrophys. J. Lett.",
    volume = "951",
    number = "1",
    pages = "L8",
    year = "2023"
}

@ARTICLE{arvanitaki2010,
  author = {Arvanitaki, Asimina and Dimopoulos, Savas and Dubovsky, Sergei and Kaloper, Nemanja and March-Russell, John},
  title = {String Axiverse},
  journal = {Phys.\ Rev.\ D},
  volume = {81},
  number = {12},
  pages = {123530},
  year = {2010},
  doi = {10.1103/PhysRevD.81.123530}
}

@article{Tiruvaskar:2025hxl,
    author = "Tiruvaskar, Shreyas and Boey, Russell and Easther, Richard and Gordon, Chris",
    title = "{Ultralight Dark Matter Constraints from NanoHertz Gravitational Waves}",
    journal = {arXiv},
    eprint = "2512.15292",
    archivePrefix = "arXiv",
    primaryClass = "astro-ph.CO",
    month = "12",
    year = "2025"
}

@ARTICLE{boey2025,
  author = {Boey, Russell and Kendall, Emily and Wang, Yourong and Easther, Richard},
  title = {Supermassive Binaries in Ultralight Dark Matter Solitons},
  journal = {Phys.\ Rev.\ D},
  volume = {112},
  number = {2},
  pages = {023510},
  year = {2025},
  doi = {10.1103/6kpp-yp75},
  eprint = {2504.16348},
  archiveprefix = {arXiv},
  primaryclass = {astro-ph.CO}
}

@ARTICLE{bullock2017,
  author = {Bullock, James S. and Boylan-Kolchin, Michael},
  title = {Small-Scale Challenges to the {$\Lambda$CDM} Paradigm},
  journal = {Annu.\ Rev.\ Astron.\ Astrophys.},
  volume = {55},
  number = {1},
  pages = {343--387},
  year = {2017},
  doi = {10.1146/annurev-astro-091916-055313}
}

@ARTICLE{chandrasekhar1943,
  author = {Chandrasekhar, S.},
  title = {Dynamical Friction.\ {I}. {General Considerations}: The Coefficient of Dynamical Friction},
  journal = {Astrophys.\ J.},
  volume = {97},
  pages = {255},
  year = {1943},
  doi = {10.1086/144517}
}

@ARTICLE{chandrasekhar1943a,
  author = {Chandrasekhar, S.},
  title = {Dynamical Friction.\ {II}. The Rate of Escape of Stars from Clusters and the Evidence for the Operation of Dynamical Friction},
  journal = {Astrophys.\ J.},
  volume = {97},
  pages = {263},
  year = {1943},
  doi = {10.1086/144518}
}

@ARTICLE{edwards2018,
  author = {Edwards, Faber and Kendall, Emily and Hotchkiss, Shaun and Easther, Richard},
  title = {{PyUltraLight}: A Pseudo-Spectral Solver for Ultralight Dark Matter Dynamics},
  journal = {J.\ Cosmol.\ Astropart.\ Phys.},
  volume = {2018},
  number = {10},
  pages = {027},
  year = {2018},
  doi = {10.1088/1475-7516/2018/10/027},
  eprint = {1807.04037},
  archiveprefix = {arXiv},
  primaryclass = {astro-ph}
}

@ARTICLE{ferreira2021,
  author = {Ferreira, Elisa G. M.},
  title = {Ultra-Light Dark Matter},
  journal = {Astron.\ Astrophys.\ Rev.},
  volume = {29},
  number = {1},
  pages = {7},
  year = {2021},
  doi = {10.1007/s00159-021-00135-6}
}

@ARTICLE{hui2017,
  author = {Hui, Lam and Ostriker, Jeremiah P. and Tremaine, Scott and Witten, Edward},
  title = {Ultralight Scalars as Cosmological Dark Matter},
  journal = {Phys.\ Rev.\ D},
  volume = {95},
  number = {4},
  pages = {043541},
  year = {2017},
  doi = {10.1103/PhysRevD.95.043541}
}

@article{Gosenca:2023yjc,
    author = "Gosenca, Mateja and Eberhardt, Andrew and Wang, Yourong and Eggemeier, Benedikt and Kendall, Emily and Zagorac, J. Luna and Easther, Richard",
    title = "{Multifield ultralight dark matter}",
    eprint = "2301.07114",
    archivePrefix = "arXiv",
    primaryClass = "astro-ph.CO",
    doi = "10.1103/PhysRevD.107.083014",
    journal = "Phys. Rev. D",
    volume = "107",
    number = "8",
    pages = "083014",
    year = "2023"
}

@article{Schwabe:2020eac,
    author = "Schwabe, Bodo and Gosenca, Mateja and Behrens, Christoph and Niemeyer, Jens C. and Easther, Richard",
    title = "{Simulating mixed fuzzy and cold dark matter}",
    eprint = "2007.08256",
    archivePrefix = "arXiv",
    primaryClass = "astro-ph.CO",
    doi = "10.1103/PhysRevD.102.083518",
    journal = "Phys. Rev. D",
    volume = "102",
    number = "8",
    pages = "083518",
    year = "2020"
}

@article{Tellez-Tovar:2021mge,
    author = "T{\'e}llez-Tovar, L. O. and Matos, Tonatiuh and V{\'a}zquez, J. Alberto",
    title = "{Cosmological constraints on the multiscalar field dark matter model}",
    eprint = "2112.09337",
    archivePrefix = "arXiv",
    primaryClass = "astro-ph.CO",
    doi = "10.1103/PhysRevD.106.123501",
    journal = "Phys. Rev. D",
    volume = "106",
    number = "12",
    pages = "123501",
    year = "2022"
    }

@article{Guo:2020tla,
    author = "Guo, Huai-Ke and Sinha, Kuver and Sun, Chen and Swaim, Joshua and Vagie, Daniel",
    title = "{Two-scalar Bose-Einstein condensates: from stars to galaxies}",
    eprint = "2010.15977",
    archivePrefix = "arXiv",
    primaryClass = "astro-ph.CO",
    doi = "10.1088/1475-7516/2021/10/028",
    journal = "JCAP",
    volume = "10",
    number = "10",
    pages = "028",
    year = "2021"
}

@article{Amin:2022pzv,
    author = "Amin, Mustafa A. and Jain, Mudit and Karur, Rohith and Mocz, Philip",
    title = "{Small-scale structure in vector dark matter}",
    eprint = "2203.11935",
    archivePrefix = "arXiv",
    primaryClass = "astro-ph.CO",
    doi = "10.1088/1475-7516/2022/08/014",
    journal = "JCAP",
    volume = "08",
    number = "08",
    pages = "014",
    year = "2022"
}

@ARTICLE{koo2024,
  author = {Koo, Hyeonmo and Bak, Dongsu and Park, Inkyu and Hong, Sungwook E. and Lee, {Jae-Weon}},
  title = {Final Parsec Problem of Black Hole Mergers and Ultralight Dark Matter},
  journal = {Phys.\ Lett.\ B},
  volume = {856},
  pages = {138908},
  year = {2024},
  doi = {10.1016/j.physletb.2024.138908}
}

@ARTICLE{marsh2016,
  author = {Marsh, David J.~E.},
  title = {Axion Cosmology},
  journal = {Phys.\ Rep.},
  volume = {643},
  pages = {1--79},
  year = {2016},
  doi = {10.1016/j.physrep.2016.06.005}
}

@ARTICLE{wang2022,
  author = {Wang, Yourong and Easther, Richard},
  title = {Dynamical Friction from Ultralight Dark Matter},
  journal = {Phys.\ Rev.\ D},
  volume = {105},
  number = {6},
  pages = {063523},
  year = {2022},
  doi = {10.1103/PhysRevD.105.063523}
}

@article{Zagorac:2022xic,
    author = "Zagorac, J. Luna and Kendall, Emily and Padmanabhan, Nikhil and Easther, Richard",
    title = "{Soliton formation and the core-halo mass relation: An eigenstate perspective}",
    eprint = "2212.09349",
    archivePrefix = "arXiv",
    primaryClass = "astro-ph.CO",
    doi = "10.1103/PhysRevD.107.083513",
    journal = "Phys. Rev. D",
    volume = "107",
    number = "8",
    pages = "083513",
    year = "2023"
}

@ARTICLE{zagorac2022,
  author = {Zagorac, J. Luna and Sands, Isabel and Padmanabhan, Nikhil and Easther, Richard},
  title = {Schr\"odinger-Poisson Solitons: Perturbation Theory},
  journal = {Phys.\ Rev.\ D},
  volume = {105},
  number = {10},
  pages = {103506},
  year = {2022},
  doi = {10.1103/PhysRevD.105.103506}
}

@ARTICLE{schive2014,
  author = {Hsi-Yu Schive and Tzihong Chiueh and Tom Broadhurst},
  title = {Cosmic Structure as the Quantum Interference of a Coherent Dark Wave},
  journal = {Nat.\ Phys.},
  volume = {10},
  number = {7},
  pages = {496--499},
  year = {2014},
  doi = {10.1038/nphys2996},
  eprint = {1406.6586},
  archiveprefix = {arXiv},
  primaryclass = {astro-ph.GA}
}

@ARTICLE{boey2024,
  author = {Russell Boey and Yourong Wang and Emily Kendall and Richard Easther},
  title = {Dynamical friction and black holes in ultralight dark matter solitons},
  journal = {Phys.\ Rev.\ D},
  volume = {109},
  number = {10},
  pages = {103526},
  year = {2024},
  doi = {10.1103/PhysRevD.109.103526},
  eprint = {2403.09038},
  archiveprefix = {arXiv},
  primaryclass = {astro-ph.CO}
}

@article{weinberg1989,
  author = {Weinberg, Martin D.},
  title = {Self-gravitating response of a spherical galaxy to sinking satellites},
  journal = {Mon. Not. R. Astron. Soc.},
  volume = {239},
  pages = {549--569},
  year = {1989},
  doi = {10.1093/mnras/239.2.549}
}

@article{weinberg1994,
  author = {Weinberg, Martin D.},
  title = {Weakly Damped Modes in Star Clusters and Galaxies},
  journal = {Astrophys.\ J.},
  volume = {421},
  number = {1},
  pages = {481--490},
  year = {1994},
  doi = {10.1086/173669}
}

@article{weinberg2023,
  author = {Weinberg, Martin D.},
  title = {New dipole instabilities in spherical stellar systems},
  journal = {Mon. Not. R. Astron. Soc.},
  volume = {525},
  pages = {4962--4975},
  year = {2023},
  doi = {10.1093/mnras/stad2591},
  eprint = {2212.02576},
  archiveprefix = {arXiv},
  primaryclass = {astro-ph.GA}
}

@article{nanograv2023,
  author = {{Agazie}, Gabriella and others},
  collaboration = {NANOGrav},
  title = {The {NANOGrav} 15 yr Data Set: Search for Signals from New Physics},
  journal = {Astrophys. J. Lett.},
  volume = {951},
  pages = {L11},
  year = {2023},
  doi = {10.3847/2041-8213/acdc91},
  eprint = {2306.16219},
  archiveprefix = {arXiv},
  primaryclass = {astro-ph.HE}
}

@article{afzal2023,
  author = {Afzal, Adeela and others},
  collaboration = {NANOGrav},
  title = {The {NANOGrav} 15 yr Data Set: Evidence for a Gravitational-wave Background},
  journal = {Astrophys. J. Lett.},
  volume = {951},
  pages = {L8},
  year = {2023},
  doi = {10.3847/2041-8213/acdac6},
  eprint = {2306.16213},
  archiveprefix = {arXiv},
  primaryclass = {astro-ph.HE}
}

@article{duque2024,
  author = {Duque, Francisco and Macedo, Caio F. B. and Vicente, Rodrigo and Cardoso, Vitor},
  title = {Extreme-Mass-Ratio Inspirals in Ultralight Dark Matter},
  journal = {Phys. Rev. Lett.},
  volume = {133},
  pages = {121404},
  year = {2024},
  doi = {10.1103/PhysRevLett.133.121404},
  eprint = {2312.06767},
  archiveprefix = {arXiv},
  primaryclass = {gr-qc}
}

@article{lancaster2020,
  author = {Lancaster, Lachlan and Giovanetti, Cara and Mocz, Philip and Kahn, Yonatan and Lisanti, Mariangela and Spergel, David N.},
  title = {Dynamical friction in a fuzzy dark matter universe},
  journal = {J.\ Cosmol.\ Astropart.\ Phys.},
  volume = {2020},
  number = {01},
  pages = {001},
  year = {2020},
  doi = {10.1088/1475-7516/2020/01/001},
  eprint = {1909.06381},
  archiveprefix = {arXiv},
  primaryclass = {astro-ph.CO}
}

@article{baror2019,
  author = {Bar-Or, Ben and Fouvry, Jean-Baptiste and Tremaine, Scott},
  title = {Relaxation in a Fuzzy Dark Matter Halo},
  journal = {Astrophys.\ J.},
  volume = {871},
  number = {1},
  pages = {28},
  year = {2019},
  doi = {10.3847/1538-4357/aaf28c},
  eprint = {1809.07673},
  archiveprefix = {arXiv},
  primaryclass = {astro-ph.GA}
}

@ARTICLE{planck2020,
  author = {{Planck Collaboration} and {Aghanim}, N. and others},
  title = {Planck 2018 results. VI. Cosmological parameters},
  journal = {Astron. Astrophys.},
  volume = {641},
  pages = {A6},
  year = {2020},
  doi = {10.1051/0004-6361/201833910}
}

@ARTICLE{springel2005,
  author = {Springel, Volker and others},
  title = {Simulations of the formation, evolution and clustering of galaxies and quasars},
  journal = {Nature},
  volume = {435},
  pages = {629--636},
  year = {2005},
  doi = {10.1038/nature03597}
}

@ARTICLE{frenk2012,
  author = {Frenk, Carlos S. and White, Simon D. M.},
  title = {Dark matter and cosmic structure},
  journal = {Annalen Phys.},
  volume = {524},
  pages = {507--534},
  year = {2012},
  doi = {10.1002/andp.201200212}
}

@ARTICLE{primack2012,
  author = {Primack, Joel R.},
  title = {Triumphs and tribulations of Lambda CDM, the double dark theory},
  journal = {Annalen Phys.},
  volume = {524},
  number = {9--10},
  pages = {535--544},
  year = {2012},
  doi = {10.1002/andp.201200077}
}

@ARTICLE{moore1994,
  author = {Moore, Ben},
  title = {Evidence against dissipationless dark matter from the rotation curves of dark haloes},
  journal = {Nature},
  volume = {370},
  pages = {629},
  year = {1994},
  doi = {10.1038/370629a0}
}

@ARTICLE{deblok2010,
  author = {de Blok, W. J. G.},
  title = {The Core-Cusp Problem},
  journal = {Adv. Astron.},
  volume = {2010},
  pages = {789293},
  year = {2010},
  doi = {10.1155/2010/789293}
}

@ARTICLE{oh2011,
  author = {Oh, Se-Heon and others},
  title = {The Central Slope of Dark Matter Halos in Dwarf Galaxies: Simulations vs. THINGS},
  journal = {Astron. J.},
  volume = {141},
  pages = {193},
  year = {2011},
  doi = {10.1088/0004-637X/737/2/87}
}

@ARTICLE{boylankolchin2011,
  author = {Boylan-Kolchin, Michael and Bullock, James S. and Kaplinghat, Manoj},
  title = {Too big to fail? The puzzling darkness of massive Milky Way subhaloes},
  journal = {Mon. Not. R. Astron. Soc.},
  volume = {415},
  pages = {L40},
  year = {2011},
  doi = {10.1111/j.1745-3933.2011.01074.x}
}

@ARTICLE{spergel2000,
  author = {Spergel, David N. and Steinhardt, Paul J.},
  title = {Observational evidence for self-interacting cold dark matter},
  journal = {Phys. Rev. Lett.},
  volume = {84},
  pages = {3760},
  year = {2000},
  doi = {10.1103/PhysRevLett.84.3760}
}

@ARTICLE{feng2010,
  author = {Feng, Jonathan L.},
  title = {Dark Matter Candidates from Particle Physics and Methods of Detection},
  journal = {Annu. Rev. Astron. Astrophys.},
  volume = {48},
  pages = {495},
  year = {2010},
  doi = {10.1146/annurev-astro-082708-101659}
}

@ARTICLE{bertone2005,
  author = {Bertone, Gianfranco and Hooper, Dan and Silk, Joseph},
  title = {Particle dark matter: Evidence, candidates and constraints},
  journal = {Phys. Rep.},
  volume = {405},
  pages = {279},
  year = {2005},
  doi = {10.1016/j.physrep.2004.08.031}
}

@ARTICLE{witten1984,
  author = {Witten, Edward},
  title = {Some properties of O(32) superstrings},
  journal = {Phys. Lett. B},
  volume = {149},
  pages = {351},
  year = {1984},
  doi = {10.1016/0370-2693(84)90422-2}
}

@ARTICLE{widrow1993,
  author = {Widrow, Lawrence M. and Kaiser, Nick},
  title = {Using the Schroedinger Equation to Simulate Collisionless Matter},
  journal = {Astrophys. J. Lett.},
  volume = {416},
  pages = {L71},
  year = {1993},
  doi = {10.1086/187073}
}

@ARTICLE{chavanis2011,
  author = {Chavanis, Pierre-Henri},
  title = {Mass-radius relation of Newtonian self-gravitating Bose-Einstein condensates with short-range interactions: I. Analytical results},
  journal = {Phys. Rev. D},
  volume = {84},
  pages = {043531},
  year = {2011},
  doi = {10.1103/PhysRevD.84.043531}
}

@ARTICLE{schive2014a,
  author = {Schive, Hsi-Yu and Liao, M.-H. and Woo, T.-P. and Wong, S.-K. and Chiueh, T. and Broadhurst, T. and Hwang, W.-Y. P.},
  title = {Understanding the Core-Halo Relation of Quantum Wave Dark Matter Structures from 3D Simulations},
  journal = {Phys. Rev. Lett.},
  volume = {113},
  pages = {261302},
  year = {2014},
  doi = {10.1103/PhysRevLett.113.261302}
}

@ARTICLE{mocz2017,
  author = {Mocz, Philip and others},
  title = {Galaxy formation with BECDM - I. Turbulence and relaxation of idealized haloes},
  journal = {Mon. Not. R. Astron. Soc.},
  volume = {471},
  pages = {4559},
  year = {2017},
  doi = {10.1093/mnras/stx1887}
}

@ARTICLE{schwabe2016,
  author = {Schwabe, B. and Niemeyer, J. C. and Engels, J. F.},
  title = {Simulations of solitonic core mergers in ultralight axion dark matter cosmologies},
  journal = {Phys. Rev. D},
  volume = {94},
  pages = {043513},
  year = {2016},
  doi = {10.1103/PhysRevD.94.043513}
}

@ARTICLE{veltmaat2018,
  author = {Veltmaat, Jan and Niemeyer, Jens C. and Schwabe, Bodo},
  title = {Formation and structure of ultralight bosonic dark matter halos},
  journal = {Phys. Rev. D},
  volume = {98},
  pages = {043509},
  year = {2018},
  doi = {10.1103/PhysRevD.98.043509}
}

@ARTICLE{lin2018,
  author = {Lin, S.-C. and Schive, H.-Y. and Wong, S.-K. and Chiueh, T.},
  title = {Self-consistent construction of virialized wave dark matter halos},
  journal = {Phys. Rev. D},
  volume = {97},
  pages = {063523},
  year = {2018},
  doi = {10.1103/PhysRevD.97.063523}
}

@ARTICLE{irsic2017,
  author = {Ir\v{s}i\v{c}, Vid and others},
  title = {First Constraints on Fuzzy Dark Matter from Lyman-$\alpha$ Forest Data and Hydrodynamical Simulations},
  journal = {Phys. Rev. Lett.},
  volume = {119},
  pages = {031302},
  year = {2017},
  doi = {10.1103/PhysRevLett.119.031302}
}

@ARTICLE{rogers2021,
  author = {Rogers, Keir K. and Peiris, Hiranya V.},
  title = {Strong bound on canonical ultralight axion dark matter from the Lyman-alpha forest},
  journal = {Phys. Rev. Lett.},
  volume = {126},
  pages = {071302},
  year = {2021},
  doi = {10.1103/PhysRevD.103.043512}
}

@ARTICLE{armengaud2017,
  author = {Armengaud, Eric and others},
  title = {Constraining the mass of light bosonic dark matter using SDSS Lyman-$\alpha$ forest},
  journal = {Mon. Not. R. Astron. Soc.},
  volume = {471},
  pages = {4606},
  year = {2017},
  doi = {10.1093/mnras/stx1870}
}

@ARTICLE{dalal2022,
  author = {Dalal, Neal and Kravtsov, Andrey},
  title = {Who Killed the Dwarf Galaxy? Constraints on Fuzzy Dark Matter from Stellar Kinematics},
  journal = {Phys. Rev. D},
  volume = {106},
  pages = {063517},
  year = {2022},
  doi = {10.1103/PhysRevD.106.063517}
}

@BOOK{binneytremaine2008,
  author = {Binney, James and Tremaine, Scott},
  title = {Galactic Dynamics: Second Edition},
  publisher = {Princeton University Press},
  address = {Princeton, NJ},
  edition = {2},
  year = {2008},
  isbn = {978-0-691-13027-9}
}

@ARTICLE{ostriker1999,
  author = {Ostriker, E. C.},
  title = {Dynamical Friction in a Gaseous Medium},
  journal = {Astrophys. J.},
  volume = {513},
  pages = {252},
  year = {1999},
  doi = {10.1086/311969}
}

@ARTICLE{just2011,
  author = {Just, Andreas and Khan, F. M. and Berczik, P. and Ernst, A. and Spurzem, R.},
  title = {Dynamical friction of massive objects in galactic centres},
  journal = {Mon. Not. R. Astron. Soc.},
  volume = {411},
  pages = {653},
  year = {2011},
  doi = {10.1111/j.1365-2966.2010.17711.x}
}

@ARTICLE{vicente2022,
  author = {Vicente, Rodrigo and Cardoso, Vitor},
  title = {Dynamical friction of black holes in ultralight dark matter},
  journal = {Phys. Rev. D},
  volume = {105},
  pages = {083008},
  year = {2022},
  doi = {10.1103/PhysRevD.105.083008}
}

@ARTICLE{epta2023,
  author = {Antoniadis, J. and others},
  collaboration = {EPTA Collaboration},
  title = {The second data release from the European Pulsar Timing Array III. Search for gravitational wave background},
  journal = {Astron. Astrophys.},
  volume = {678},
  pages = {A50},
  year = {2023},
  doi = {10.1051/0004-6361/202346844}
}

@ARTICLE{ppta2023,
  author = {Reardon, D. J. and others},
  collaboration = {PPTA Collaboration},
  title = {Search for an Isotropic Gravitational-wave Background with the Parkes Pulsar Timing Array},
  journal = {Astrophys. J. Lett.},
  volume = {951},
  pages = {L6},
  year = {2023},
  doi = {10.3847/2041-8213/acdd02}
}

@ARTICLE{milosavljevic2003,
  author = {Milosavljevi\'c, Milo\v{s} and Merritt, David},
  title = {The Final Parsec Problem},
  journal = {AIP Conf. Proc.},
  volume = {686},
  pages = {201},
  year = {2003},
  doi = {10.1086/378086}
}

@ARTICLE{begelman1980,
  author = {Begelman, M. C. and Blandford, R. D. and Rees, M. J.},
  title = {Massive black hole binaries in active galactic nuclei},
  journal = {Nature},
  volume = {287},
  pages = {307},
  year = {1980},
  doi = {10.1038/287307a0}
}

@ARTICLE{kelley2017,
  author = {Kelley, Luke Zoltan and others},
  title = {The gravitational wave background from massive black hole binaries in Illustris: spectral features and time to detection with pulsar timing arrays},
  journal = {Mon. Not. R. Astron. Soc.},
  volume = {471},
  pages = {4508},
  year = {2017},
  doi = {10.1093/mnras/stx1638}
}

@ARTICLE{khan2013,
  author = {Khan, Fazeel M. and others},
  title = {Mergers of supermassive black holes in galactic mergers},
  journal = {Astrophys. J.},
  volume = {773},
  pages = {100},
  year = {2013},
  doi = {10.1088/0004-637X/773/2/100}
}

@ARTICLE{vasiliev2015,
  author = {Vasiliev, Eugene and Antonini, Fabio and Merritt, David},
  title = {The Final Parsec Problem in Non-spherical Galaxies},
  journal = {Astrophys. J.},
  volume = {810},
  pages = {49},
  year = {2015},
  doi = {10.1093/mnras/stv1304}
}

@ARTICLE{gould2000,
  author = {Gould, Andrew and Rix, Hans-Walter},
  title = {Binary Black Hole Mergers from Planet-like Migrations},
  journal = {Astrophys. J. Lett.},
  volume = {532},
  pages = {L29},
  year = {2000},
  doi = {10.1086/312562}
}

@ARTICLE{tremaineweinberg1984,
  author = {Tremaine, Scott and Weinberg, Martin D.},
  title = {Dynamical friction in spherical systems},
  journal = {Mon. Not. R. Astron. Soc.},
  volume = {209},
  pages = {729},
  year = {1984},
  doi = {10.1086/162893}
}

@ARTICLE{hu2000,
  author = {Hu, Wayne and Barkana, Rennan and Gruzinov, Andrei},
  title = {Fuzzy Cold Dark Matter: The Wave Properties of Ultralight Particles},
  journal = {Phys. Rev. Lett.},
  volume = {85},
  pages = {1158--1161},
  year = {2000},
  doi = {10.1103/PhysRevLett.85.1158}
}

@ARTICLE{weinberg2015,
  author = {Weinberg, David H. and Bullock, James S. and Governato, Fabio and Kuzio de Naray, Rachel and Peter, Annika H. G.},
  title = {Cold dark matter: controversies on small scales},
  journal = {Proc. Natl. Acad. Sci. U.S.A.},
  volume = {112},
  number = {40},
  pages = {12249--12255},
  year = {2015},
  doi = {10.1073/pnas.1308716112}
}

@ARTICLE{svrcek2006,
  author = {Svrcek, Peter and Witten, Edward},
  title = {Axions in string theory},
  journal = {J. High Energy Phys.},
  volume = {06},
  number = {06},
  pages = {051},
  year = {2006},
  doi = {10.1088/1126-6708/2006/06/051}
}

@ARTICLE{cardoso2022,
  author = {Cardoso, Vitor and others},
  title = {Gravitational-wave emission from the motion of black holes in ultralight dark matter},
  journal = {J. Cosmol. Astropart. Phys.},
  volume = {07},
  number = {07},
  pages = {048},
  year = {2022},
  doi = {10.1088/1475-7516/2022/07/048}
}

@article{may2025,
    author = "May, Simon and Dalal, Neal and Kravtsov, Andrey",
    title = "{Updated bounds on ultra-light dark matter from the tiniest galaxies}",
    journal = {arXiv preprint},
    eprint = "2509.02781",
    primaryClass = "astro-ph.CO",
    month = "9",
    year = "2025"
}

@article{eberhardt2025,
    author = "Eberhardt, Andrew and Gosenca, Mateja and Hui, Lam",
    title = "{Heating and scattering of stellar distributions by ultralight dark matter}",
    journal = {arXiv preprint},
    eprint = "2510.17079",
    primaryClass = "astro-ph.CO",
    month = "10",
    year = "2025"
}

@misc{eberhardt2025review,
      title={Ultralight fuzzy dark matter review}, 
      author={Andrew Eberhardt and Elisa G. M. Ferreira},
      year={2025},
      eprint={2507.00705},
      archivePrefix={arXiv},
      primaryClass={astro-ph.CO},
      url={https://arxiv.org/abs/2507.00705}, 
}

@article{Barmak2024,
  title = {Supermassive black hole binaries in ultralight dark matter},
  author = {Bromley, Benjamin C. and Sandick, Pearl and Shams Es Haghi, Barmak},
  journal = {Phys. Rev. D},
  volume = {110},
  pages = {023517},
  year = {2024},
  doi = {10.1103/PhysRevD.110.023517}
}

@article{li2021,
  author       = {Li, Xinyu and Hui, Lam and Yavetz, Tomer D.},
  title        = {Oscillations and Random Walk of the Soliton Core in a Fuzzy Dark Matter Halo},
  journal      = {Phys. Rev. D},
  volume       = {103},
  pages        = {023508},
  year         = {2021},
  doi          = {10.1103/PhysRevD.103.023508},
  eprint       = {2011.11416},
  archivePrefix= {arXiv},
  primaryClass = {astro-ph.CO}
}

@article{widmark2024,
  author       = {Widmark, Axel and Yavetz, Tomer D. and Li, Xinyu},
  title        = {Fuzzy dark matter dynamics in tidally perturbed dwarf spheroidal galaxy satellites},
  journal      = {JCAP},
  volume       = {03},
  number = {03},
  pages        = {052},
  year         = {2024},
  doi          = {10.1088/1475-7516/2024/03/052},
  eprint       = {2309.00039},
  archivePrefix= {arXiv},
  primaryClass = {astro-ph.GA}
}

@article{yavetz2022,
  author       = {Yavetz, Tomer D. and Li, Xinyu and Hui, Lam},
  title        = {Construction of Wave Dark Matter Halos: Numerical Algorithm and Analytical Constraints},
  journal      = {Phys. Rev. D},
  volume       = {105},
  pages        = {023512},
  year         = {2022},
  doi          = {10.1103/PhysRevD.105.023512},
  eprint       = {2109.06125},
  archivePrefix= {arXiv},
  primaryClass = {astro-ph.CO}
}

\end{document}